\documentclass[submission]{SciPost}

\hypersetup{
    colorlinks,
    linkcolor={red!50!black},
    citecolor={blue!50!black},
    urlcolor={blue!80!black}
}

\usepackage[bitstream-charter]{mathdesign}
\DeclareSymbolFont{usualmathcal}{OMS}{cmsy}{m}{n}
\DeclareSymbolFontAlphabet{\mathcal}{usualmathcal}

\usepackage{hyperref}
\usepackage{xspace}
\usepackage{mathtools}
\usepackage{wasysym}
\usepackage{stmaryrd}
\usepackage{marvosym}
\usepackage{stackengine}
\usepackage{setspace}
\usepackage{esvect}
\usepackage{comment}
\usepackage{color}
\usepackage{ifthen}
\newcommand{\beq}{\begin{equation}}
\newcommand{\eeq}{\end{equation}}
\newcommand{\bea}{\begin{eqnarray}}
\newcommand{\eea}{\end{eqnarray}}

\def\nn{\nonumber\\}
\def\fr#1{(\ref{#1})}

\newcommand{\LT}{lifted \TASEP}
\newcommand{\LLT}{Lifted \TASEP}
\newcommand{\GLT}{GL-\TASEP}
\newcommand{\LUT}{lifted-\TASEP\xspace}

\newcommand{\ZERO}{\fbox{\phantom{$\bullet$}}}
\newcommand{\ONE}{\fbox{$\bullet$}}

\newcommand{\TWO}{\fbox{$\mathrlap{\mkern-4mu \rightarrow}\bullet$}}

\newcommand{\alphacrit}{\alpha_\text{crit}}

\newcommand{\TASEP}{\protect \textsc{Tasep}\xspace}

\newcommand{\SSEP}{\textsc{Ssep}\xspace}

\def\EV{E}
\def\sfix#1{\texorpdfstring{#1}{Lg}}

\newcommand{\Lcont}{\LCAL}

\newcommand{\pfact}{p^{\text{fact}}}
\newcommand{\REF}[2][]{
	\ifthenelse{\equal {#1} {}}{Ref.~\cite{#2}}{Ref.~\cite[#1]{#2}}}
\renewcommand{\emph}[1]{\textit{#1}}
\newcommand{\GLTASEP}{GL-\TASEP}
\newcommand{\rightEV}[1]{\psi^R({#1})}

\newcommand{\Lfrak}{{\mathfrak L}}
\newcommand{\afrak}{{\mathfrak a}}
\newcommand\subcap[1]{{(#1):}}

\newcommand{\SET}[1]{\{#1\}}

\newcommand{\eq}[1]{Eq.~(\ref{#1})}
\newcommand{\eqq}[1]{Equation~(\ref{#1})}
\newcommand{\eqtwo}[2]{Eqs~(\ref{#1}) and~(\ref{#2})}
\newcommand{\fig}[1]{Fig.~\ref{#1}}

\newcommand{\quot}[1]{``#1''}

\newcommand{\sect}[1]{Section~\ref{#1}}

\newcommand{\LCAL}{\mathcal{L}}  %  mathcal
\newcommand{\OCAL}{\mathcal{O}}  %  mathcal
\newcommand{\alphabar}{\overline{\alpha}}  %  overbar 
\newcommand{\pbar}{\overline{p}}  %  overbar
\newcommand{\expa}[1]{\mathrm{e}^{#1}}   % high exponential groupings a
\newcommand{\expb}[1]{\exp \glb #1 \grb} % low exponential with groupings b
\newcommand{\logc}[2][]{\log^{#1} \glc #2 \grc}  % log-brace,  with - []
\newcommand{\minb}[2][]{\min^{#1} \glb #2 \grb}  % min-brace,  with - ()
\newcommand{\minc}[2][]{\min^{#1} \glc #2 \grc}  % min-brace,  with - []
\newcommand{\glb}{\left(}  % ' group left b' 
\newcommand{\grb}{\right)}  % ' group right b' 
\newcommand{\glc}{\left[}  % ' group left c'
\newcommand{\grc}{\right]}  % ' group right c' 
\newcommand{\TO}{,\ldots,}

\newcommand{\del}{\delta}
\newcommand{\mean}[1]{\left\langle #1 \right\rangle}

\newcommand\bigOb[1]{\ensuremath{\OCAL\glb #1 \grb}}

\usepackage{float}
\begin{document}
\newfloat{algorithm}{ht}{loa}
\floatname{algorithm}{Algorithm }

\begin{center}{\Large \textbf{Lifted TASEP: long-time dynamics,
      generalizations, and continuum limit}}
\end{center}

\begin{center}\textbf{
Fabian H.L.  Essler\textsuperscript{1$\star$},
Jeanne Gipouloux\textsuperscript{2} and
Werner Krauth\textsuperscript{2,1,3$\dagger$}
}\end{center}

\begin{center}
{\bf 1}
Rudolf Peierls Centre for Theoretical Physics, Clarendon
Laboratory, Oxford OX1 3PU, UK
\\
{\bf 2}
Laboratoire de Physique de l’Ecole normale sup\'erieure, ENS,
Universit\'e PSL, CNRS, Sorbonne Universit\'e, Universit\'e de Paris Cit\'e,
24 rue Lhomond, 75005 Paris, France
\\
{\bf 3} Simons Center for Computational Physical Chemistry,
New York University, New York (NY), USA

%%%%%%%%%% END TODO: AFFILIATIONS
%%%%%%%%%% TODO: EMAIL
% Provide email address of corresponding author(s)
$\star$ \href{mailto:email1}{\small fab@thphys.ox.ac.uk}\,,\quad
$\dagger$ \href{mailto:email2}{\small werner.krauth@ens.fr}
\\[\baselineskip]
%%%%%%%%%% END TODO: EMAIL
\end{center}

\section*{{Abstract}}
{\boldmath\bf
%%%%%%%%%% TODO: ABSTRACT
% Write your abstract here.
We investigate the \LT and its generalization, the \GLT. We
analyze the spectral properties of the transition matrix of the \LT
using its Bethe ansatz solution, and use them to determine the scaling
of the relaxation time (the inverse spectral gap) with particle
number. The observed scaling with particle number was previously
found to disagree with Monte Carlo simulations of the equilibrium
autocorrelation times of the structure factor and of other large-scale
density correlators for a particular value of the pullback
$\alpha_{\rm crit}$. We explain this discrepancy. We then construct
the continuum limit of the \LT, which remains integrable, 
and connect it to the event-chain Monte Carlo algorithm. The critical
pullback $\alpha_{\rm crit}$ then equals the system pressure. We
generalize the \LT to a large class of nearest-neighbour interactions,
which lead to stationary states characterized by non-trivial Boltzmann
distributions. By tuning the pullback parameter in the \GLT to a
particular value we can again achieve a polynomial speedup in the
time required to converge to the steady state. We comment on the
possible integrability of the \GLT.

%%%%%%%%%% END TODO: ABSTRACT
}

\vspace{\baselineskip}

%\tableofcontents
%%%%%%%%%%%%%%%%%%%%%%%%%%%%%%
\section{Introduction}
\label{sec:Introduction} 
%%%%%%%%%%%%%%%%%%%%%%%%%%%%%%
In recent years, Monte Carlo algorithms based on non-reversible Markov chains
have received a growing amount of attention. In several important
applications~\cite{Bernard2011,Kampmann2021,Klement2019}, they empirically
outperform reversible Monte Carlo algorithms built on the detailed-balance
condition. The message of these works is that breaking reversibility
can improve on the slow diffusive exploration of high-dimensional
sample spaces.

The theoretical analysis of lifted \cite{Diaconis2000,Chen1999}
non-reversible Markov chains is involved because typically the
spectrum of the transition matrix is complex valued and (left or right)
eigenvectors do not form an orthonormal
basis~\cite{Levin2008,Krauth2021eventchain}.  
In order to bridge the gap between exactly solved lifted single-particle
models~\cite{Diaconis2000,Chen1999} and real-life
applications~\cite{Hoellmer2024}, the study of non-reversible lifted
Markov chains for interacting many-particle systems in one spatial
dimension was initiated in Refs~\cite{KapferKrauth2016,Lei2018_OneD,Lei2019}. In
continuum systems, this leads to implementations of the
event-chain Monte Carlo algorithm~\cite{Bernard2009,Michel2014JCP}, while
for lattice systems this connects lifted Markov chains to the vast
literature on exactly solvable (both reversible and irreversible)
Markov-chain models such as the
asymmetric simple exclusion
process\cite{Spitzer1970,Gwa1992six,Gwa1992bethe,Kim1995Bethe,Derrida_1999,
golinelli2004bethe,deGier2005Bethe,Lee_2006,de2006exact,de2008slowest,Simon09,
deGierEssler11,mallick2011some,crampe2011matrix,Lazarescu_2014,Prolhac_2016}.

In a recent work~\cite{essler2024lifted}, two of us proposed the \LT as a
paradigm for
non-reversible \emph{lifted} Markov chains in one-dimensional particle
systems. The model considers $N$ hard-sphere particles on an $L$-site lattice with periodic
boundary conditions with only a single particle being active. It carries a pointer which
allows it to move in forward direction or to undergo a collision. In a second part of
the move, the pointer itself moves to the nearest neighbour to the left, which
becomes the new active particle:
\begin{align}
\underbrace{ \ONE \TWO \ZERO \ONE}_{x_t}
&\rightarrow
\ONE \ZERO \TWO  \ONE
\rightarrow
\underbrace{
\begin{cases}
 \TWO \ZERO \ONE \ONE & \alpha\\
 \ONE \ZERO \TWO  \ONE &  1 - \alpha
\end{cases}}_{x_{t+1}}
\label{equ:LTASEP1}
 \\
\underbrace{\ONE \TWO \ONE \ZERO}_{x_t}
&\rightarrow
 \ONE \ONE \TWO  \ZERO
    \rightarrow
\underbrace{
\begin{cases}
 \ONE \TWO \ONE \ZERO  &\alpha \\
 \ONE \ONE \TWO \ZERO  & 1 - \alpha
\end{cases}}_{x_{t+1}} \ .
\label{equ:LTASEP2}
\end{align}
For all $0<\alpha <1$,  the \LT is irreducible and
aperiodic~\cite{Levin2008,essler2024lifted}, and the
$\alpha$-independent steady state is the equal-probability mixture of all
possible lifted configurations (which are characterized by the
positions of particles and the pointer). In
Ref.~\cite{essler2024lifted} it was shown that the \LT is exactly
solvable by means of Bethe-like ansatz, which was then used to
establish bounds of the scaling of the spectral gap $\Delta$ with particle
number. The results reported in Ref.~\cite{essler2024lifted} suggested
that, for $N, L \to \infty$ with $N/L$ fixed, one has
\beq
\Delta={\rm Re}\ln[E^*]\leq \begin{cases}
  {\rm const}\ N^{-5/2} & \text{ if } \alpha\neq\alphacrit\ ,\\
  {\rm const}\ N^{-2} & \text{ if } \alpha=\alphacrit\ ,\\
  \end{cases}
\eeq
Here $\alphacrit=N/L$ and $E^* \neq 1$ is the eigenvalue of the transition
matrix with magnitude closest to one. These results were complemented
by numerical simulations of the integrated autocorrelation time of the
equilibrium structure factor $\tau_{\rm IAC}$, which were compatible with
\beq
\tau_{\rm IAC}\sim\begin{cases}
N^{5/2} & \text{if }\alpha\neq\alphacrit\ ,\nn
N^{3/2} & \text{if }\alpha=\alphacrit\ .
\end{cases}
\eeq
Interestingly, at the special value $\alpha=\alphacrit$, the \LT
displays a polynomial speedup in its approach to the steady
state, on top of the $\sim N^{1/2}$ speedup that the generic-$\alpha$ case
achieves compared to the \SSEP (the symmetric simple exclusion
process, in other words the Metropolis algorithm). However, as
noted in \REF{essler2024lifted}, the scalings of $\tau_{\rm IAC}$ and
$\Delta$ do not agree at $\alpha=\alphacrit$. One of the main aims of
the present work is to resolve this discrepancy.

The outline of this work is as follows. In
\sect{sec:relaxationtime}, we expand on the analyses of both $\Delta$
and $\tau_{\rm IAC}$ as functions of particle number $N$ and the
pullback parameter $\alpha$. By numerically solving the Bethe
equations for particular families of eigenstates of the transition
matrix and considering particle numbers up to $N\sim 500$ (compared to
$N\sim 250$ in \REF{essler2024lifted}) we are able to clearly exhibit the
crossover (as a function of $\alpha$) between the asymptotic
$N^{-5/2}$ scaling of $\Delta$ at $\alpha\neq\alphacrit$ and the
$N^{-2}$ scaling at $\alpha=\alphacrit$. We observe the same
crossover behaviour in Monte Carlo simulations of the integrated
autocorrelation time of the structure factor.
In \sect{sec:resolution} we turn to the discrepancy (see above)
between the scaling behaviours in $\tau_{\rm IAC}$ and $\Delta$ at
$\alpha=\alphacrit$. By carefully keeping track of the translational
invariance of the problem we identify which observables are a priori
sensitive to the eigenvector of the transition matrix that gives rise
to the scaling of $\Delta$. We then show that for small particle
numbers the contribution of the eigenvector of interest to dynamical
susceptibilities is too small to be observed in Monte Carlo
simulations. We propose that this smallness of the relevant matrix
elements, combined with a small number of eigenvectors whose
eigenvalues scale as $N^{-2}$, makes it essentially impossible to
detect the asymptotic relaxation time numerically. In
\sect{sec:contlim}, we construct a continuum limit (in both space and
time) of the \LT that remains integrable and derive the Bethe ansatz
equations that determine the eigenvalues of the transition matrix. We 
discuss the equivalence of this continuum process with the hard-sphere
event-chain Monte Carlo algorithm. In particular, the pullback
$\alpha$ is related to the pressure, as discussed previously, and the
critical pullback $\alphacrit$ is seen to correspond to a vanishing
pressure \cite{Lei2019}.
In \sect{sec:GeneralizedLiftedTASEP}, we generalize the \LT to a wide
class of nearest-neighbour interactions. The \GLTASEP by construction
provides a lifted Markov chain whose stationary state is the targeted
Boltzmann distribution of interest. We present some preliminary
results on the possible integrability of the \GLTASEP.
Finally, \sect{sec:Conclusions} contains our conclusions.

%%%%%%%%%%%%%%%%%%%%%%%%%%%%%%%%%%%%%%%%%%%%%%%%%
\section{\LLT: relaxation and autocorrelation times}
\label{sec:relaxationtime}
%%%%%%%%%%%%%%%%%%%%%%%%%%%%%%%%%%%%%%%%%%%%%%%%%

%%%%%%%%%%%%%%%%%%%%%%%%%%%%%%%%
\subsection{Monte Carlo computations of autocorrelation times}
%%%%%%%%%%%%%%%%%%%%%%%%%%%%%%%%
In order to determine the scaling of the relaxation time with particle
number we have carried out extensive Monte Carlo simulations of
autocorrelation functions. These start from a random configuration
$x_0$ and under time evolution give rise to a trajectory $\SET{x_0,
x_1, x_2, \dots}$. For individual trajectories
$x_t=\{j_1(t),\dots,j_N(t); p(t)\}$ where 
$j_n(t)$ and $p(t)$ denote respectively the particle and pointer
positions we determine time-dependent observables $f_t=f(x_t)$, and
use these to determine autocorrelation functions
\begin{equation}
 C(t) = \mean{f_s f_{s+t}} - \mu^2.
\end{equation}
Here, $\mean{\ldots}$ denotes a sample average and $\mu=\mean{f}$. In
practice, it is convenient to focus on integrated autocorrelation times of
normalized autocorrelation functions \cite{Sokal1997}, which are defined as
\begin{equation}
\tau_f = \frac{1}{2} \sum_{t = -\infty}^{\infty} \frac{C(t)}{C(0)}.
\end{equation}
In the following, we present results for the structure factor $f_t =
|S(2\pi/L,t)|^2$, defined as
\begin{align}
|S(q,t)|^2 &= \frac{1}{N} \Big| \sum_{r=1}^L\expa{i q r}\rho(r,t)\Big|^2 ,\
\rho(r,t)=\sum_{n=1}^N\delta_{r,j_n(t)}\ .
\label{equ:StructureFactor}
\end{align}
This observable is sensitive to long-range density fluctuations, which are
expected to relax slowly in equilibrium. We have analyzed a number of
other observables in order to verify that the relaxational behaviour
seen for the structure factor is generic. Examples are
\beq
S(q,t)=\frac{1}{\sqrt{L}}\sum_{r=1}^L e^{iqr}\rho(r,t)\ ,\qquad
{\cal O}_1(q,d)=\sum_{r,r'=1}^L
e^{iqr}\rho(r,t)\rho(r',t)\delta_{|r-r'|,d}\ .
\eeq

\begin{figure}[ht]
\centering
\includegraphics[width=0.60\linewidth]{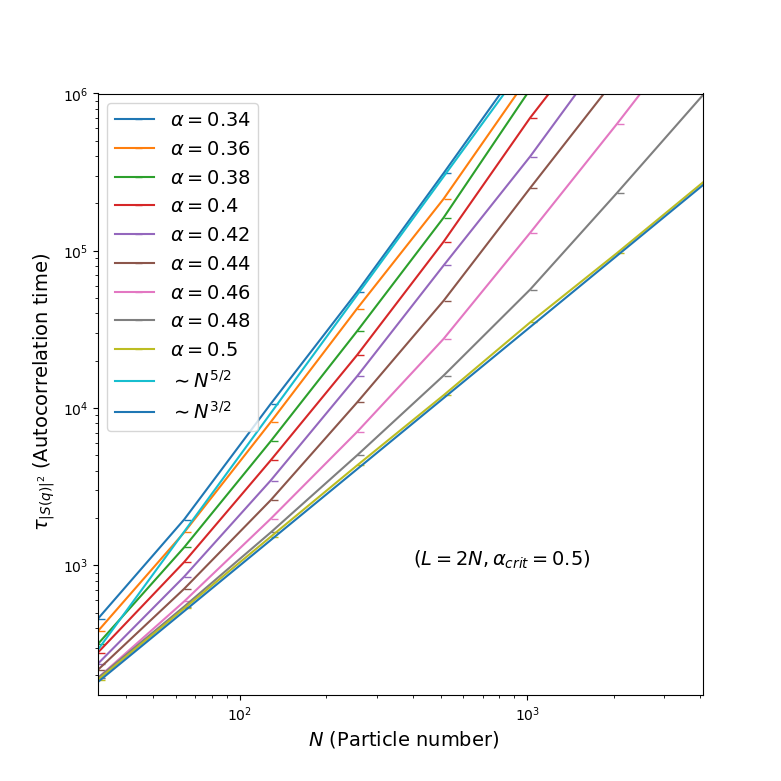}
\caption{Autocorrelation functions of the
structure factor at $L = 2N$ (with $\alphacrit = 0.5$), for different values of
the pullback $\alpha$. For $\alpha  < \alphacrit$, the asymptotic scaling is
as $\sim N^{5/2}$, but for  $\alpha \lesssim \alphacrit$, this is reached only
for very large $N$. The scalings $\sim N^{3/2}$ and $\sim N^{5/2}$ are indicated
through straight lines.}
\label{fig:AutocorrelationAlpha}
\end{figure}

Our equilibrium Markov-chain Monte Carlo simulations, presented in
\REF{essler2024lifted}, show quite clearly that
\begin{equation}
  \tau_{\rm IAC}\equiv
  \tau_{|S(q)|^2}\sim
\begin{cases}
N^{5/2} & \text{for $\alpha\neq\alphacrit$}, \\
N^{3/2} & \text{for $\alpha =\alphacrit$}.
\end{cases}
\end{equation}
For $\alpha \lesssim \alphacrit$, the asymptotic $N^{5/2}$ scaling is
reached only for large system sizes (see \fig{fig:AutocorrelationAlpha}).

%%%%%%%%%%%%%%%%%%%%%%%%%%%%%%%%
\subsection{Bethe ansatz solution of the \LT}
%%%%%%%%%%%%%%%%%%%%%%%%%%%%%%%%
Configurations in the \LT are labelled by the positions $j_1< \dots < j_N$
of the $N$ particles and an integer $1\leq a\leq N$, which identifies
the pointer among the particles. In this basis, the eigenvectors of the
transition matrix have amplitudes 
\beq
\psi_a(\boldsymbol{j})=\sum_{Q\in S_N}A_a(Q)\prod_{j=1}^N
(z_j)^{j_{Q_j}}\ ,
\label{ansatz}
\eeq
where the rapidities $z_j$ and $E$ are solutions to the following set of
coupled equations\cite{essler2024lifted}
\begin{align}
&z_a^{L-1}=\frac{\frac{1-\alpha}{z_a} -E}{\alpha}(-1)^N\prod_{b=1}^N
\frac{E-\alpha-\frac{1-\alpha}{z_b}}{E-\alpha-\frac{1-\alpha}{z_a}}
\ ,\quad a=1,\dots,N\ ,\nn
  &\prod_{j=1}^N\left(E-\frac{1-\alpha}{z_j} \right)
  =\alpha^N\prod_{k=1}^N\frac{1}{z_k}\ . 
\label{equ:BAE}
\end{align}
Periodic boundary conditions imply that
\begin{align}
\psi_N(j_1,\dots,j_{N-1},L+1)&=\psi_1(1,j_1,\dots,j_{N-1})\ ,\nn
\psi_1(0,j_2,\dots,j_{N})&=\psi_N(j_2,\dots,j_{N},L)\ .
\label{PBC}
\end{align}
The $L$-site lattice with periodic boundary conditions is
translationally invariant and, therefore,  the eigenvectors of the
transition matrix all have definite momenta. The latter can be
worked out by considering a translation by one site, and by observing
that \eqtwo{ansatz}{PBC} imply that
\beq
\psi_a(j_1+1,\dots,j_N+1)=(\prod_{n=1}^Nz_n)\ \psi_a(j_1,\dots,j_N)\ .
\eeq
We conclude that the momentum of a left eigenstate with rapidities
$\{z_j\}$ is
\beq
P=i\sum_{n=1}^N\ln(z_n)\ .
\eeq
In the case $L=2N$, we may reparametrize the Bethe ansatz
equations using  
\begin{align}
\beta&=\frac{\alphabar}{\EV-\alpha}\ ,\quad u_a=\frac{2z_a}{\beta}-1\ ,\
\delta=\frac{\beta- \alphabar/ \alpha }{\beta+
\alphabar / \alpha}\ ,\nn
\mu&=\left(\frac{2}{\beta}\right)^L\frac{1-\alpha(1-\beta)}{2\alpha}
\prod_{b=1}^N\frac{u_b-1}{u_b+1}\ .
\end{align}
This maps \eq{equ:BAE} onto
\begin{align}
\big(1-u_a^2\big)^{\frac{L}{2}}&=-\mu(u_a+\delta)\ ,\quad
a=1,\dots,\frac{L}{2}\ ,\nn
\glc \frac{2\alpha}{1-\alpha(1-\beta)}\grc^{\frac{L}{2}}&
=\prod_{b=1}^{\frac{L}{2}}(u_b+\delta)\ .
\label{BAEu}
\end{align}
%%%%%%%%%%%%%%%%%%%%%%%%%%%%%%%%
%\subsection{Steady State}
%%%%%%%%%%%%%%%%%%%%%%%%%%%%%%%%

In terms of the solutions of the Bethe equations, the steady state has
$E=1$ and corresponds to~\cite{essler2024lifted}
\beq
z_a\longrightarrow 1\ ,\quad a=1,\dots,N.
\eeq

%%%%%%%%%%%%%%%%%%%%%%%%%%%%%%%%%%%%%%%%%%%%%%%%%%%%%%%%%%%%%%%%
\subsection{Excited states for $L=2N$ in the zero-momentum sector}
%%%%%%%%%%%%%%%%%%%%%%%%%%%%%%%%%%%%%%%%%%%%%%%%%%%%%%%%%%%%%%%%
We determine excited states by means of the following procedure. As
is usually the case in Bethe ansatz solvable models
\cite{korepin1997quantum,takahashi1999thermodynamics,essler2005one,gaudin2014bethe},
the roots characterizing the eigenstates approach a limiting
distribution in the thermodynamic limit. In most cases studied in the
literature, the root distribution can be described in terms of
(half-odd) integer numbers that follow a given pattern that can be
easily followed when $L$ is increased at fixed particle number. In
order for this to work it is essential to have a suitable
parametrization of the Bethe equations such that, upon taking
logarithms with suitably defined branch cuts, one obtains a one-to-one
correspondence between eigenstates and (half-odd) integer numbers
\cite{yang1969thermodynamics,takahashi1999thermodynamics,essler2005one}.
Given the unusual form of our Bethe equations, it is not clear how to
achieve this for the \LT. We therefore proceed in an iterative way. Let us
assume that we have constructed a family of eigenstates at fixed
density $\frac{N}{L}=\frac{1}{2}$ up to a given system size. From the
roots we can determine the corresponding sets $\{\mu_\ell\}$,
$\{\delta_\ell\}$ and $\{\beta_\ell\}$.
\begin{enumerate}
\item{} Using an extrapolation algorithm we determine
$\mu_{L'}$, $\delta_{L'}$, $\beta_{L'}$ with (even) $L'\geq L+2$ from the 
sequences $\{\mu_\ell\}$, $\{\delta_\ell\}$ and $\{\beta_\ell\}$.
\item{} We then determine the $L'$ roots of the polynomial equation
  \fr{BAEu}. Out of these we select the $L'/2$ roots that most closely
  follow the pattern set out by the solution $\{u_1,\dots,u_{L/2}\}$
  for system size $L$. This gives us a starting point for solving the
  Bethe equations for system size $L'$.
\item{} We numerically solve the Bethe equations \fr{BAEu} for system
  size $L'$. 
\end{enumerate}
The starting point for our procedure is established by numerically
solving the Bethe equations for small system sizes $L=10,12,14$ and
identifying eigenstates that belong to the same family ``by hand''.

For $\alpha\neq 1/2$ the eigenvalue obtained for large sizes $L$ is
then fitted to the functional form 
\beq
E(L)=c_1L^{-5/2}+c_2L^{-3}+c_3L^{-7/2}\ .
\label{eofl}
\eeq
In order to ensure that we are ``following'' the correct state by our
iterative procedure, we determine a set of integers obtained by taking
the logarithm of the Bethe equations 
\beq
2\pi iI_j=\frac{L}{2}\ln\big(u_j-1\big)
+\frac{L}{2}\ln\big(-u_j-1\big)-\ln(\mu)-
\ln\big(u_j+\del\big)\ ,\quad j=1,\dots,\frac{L}{2}\ .
\label{ints}
\eeq
In practice, we choose a variety of different branch cuts, which gives
rise to different definitions of the $I_j$. Importantly, the
definition \fr{ints} does not rule out that the same integer $I_j$
occurs more than once, but this is not relevant for our purposes. In
the following, we focus on three classes of excited states:
\begin{itemize}
\item{} {\bf State 1:} 

The first state has zero momentum and
\beq
E,\mu,\delta,\beta\in\mathbb{R}\ .
\eeq
For $L=4n+2$ with $n$ a positive integer, it is associated with a
sequence of $I_j$ \fr{ints} of the form 
\beq
-\frac{L-2}{4},-\frac{L-2}{4}+1,\dots,-2,0,0,0,2,3\dots,\frac{L-2}{4}\ .
\label{ints1}
\eeq
We have considered this state for $\alpha\leq\frac{1}{2}$.

\item{} {\bf State 2:}

The second state has zero momentum and
\beq
E,\mu,\delta,\beta\in\mathbb{R}\ .
\eeq
For $L=4n+2$ with $n$ a positive integer, it is associated with a
sequence of $I_j$ of \eq{ints} with the form
\beq
-\frac{L-2}{4},-\frac{L-2}{4}+1,\dots,\frac{L-2}{4}\ .
\label{ints2}
\eeq
We have considered this state for $\alpha>\frac{1}{2}$. We note that
results for the same state at $\alpha=0.9$ were already reported in
\REF{essler2024lifted}.

\item{} {\bf State 3:}  

The third state has momentum $P=\frac{2\pi}{L}$ \footnote{There is a
corresponding state with $P=-\frac{2\pi}{L}$ that is related by complex
conjugation of the roots and of $E$.} and
\beq
E,\mu,\delta,\beta\in\mathbb{C}\ .
\eeq
For $L=4n+2$ with $n$ a positive integer, it is associated with a
sequence of $I_j$ of \eq{ints} of the form
\beq
-\frac{L-2}{4},-\frac{L-2}{4}+1,\dots,\frac{L-2}{4}\ .
\label{ints3}
\eeq
This state is of particular interest at $\alpha=1/2$.
\end{itemize}

%%%%%%%%%%%%%%%%%%%%%%%%%%%%%%%%%%%%%%%%%%%%%%%%%
\subsubsection{State 1 at \sfix{$\alpha=0.1$}}
\begin{figure}[ht]
\centering
\includegraphics[width=0.45\linewidth]{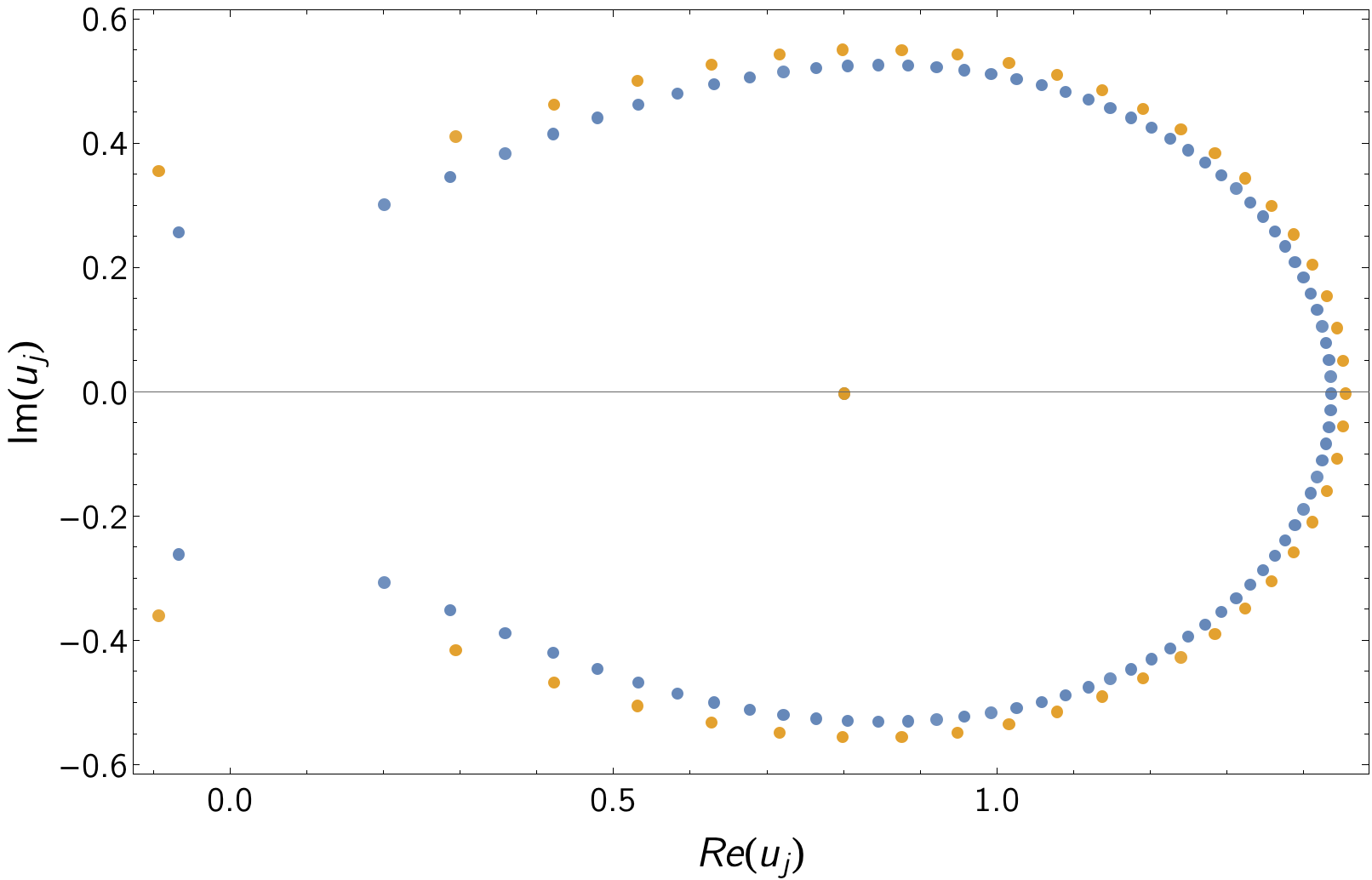}
\qquad
\includegraphics[width=0.45\linewidth]{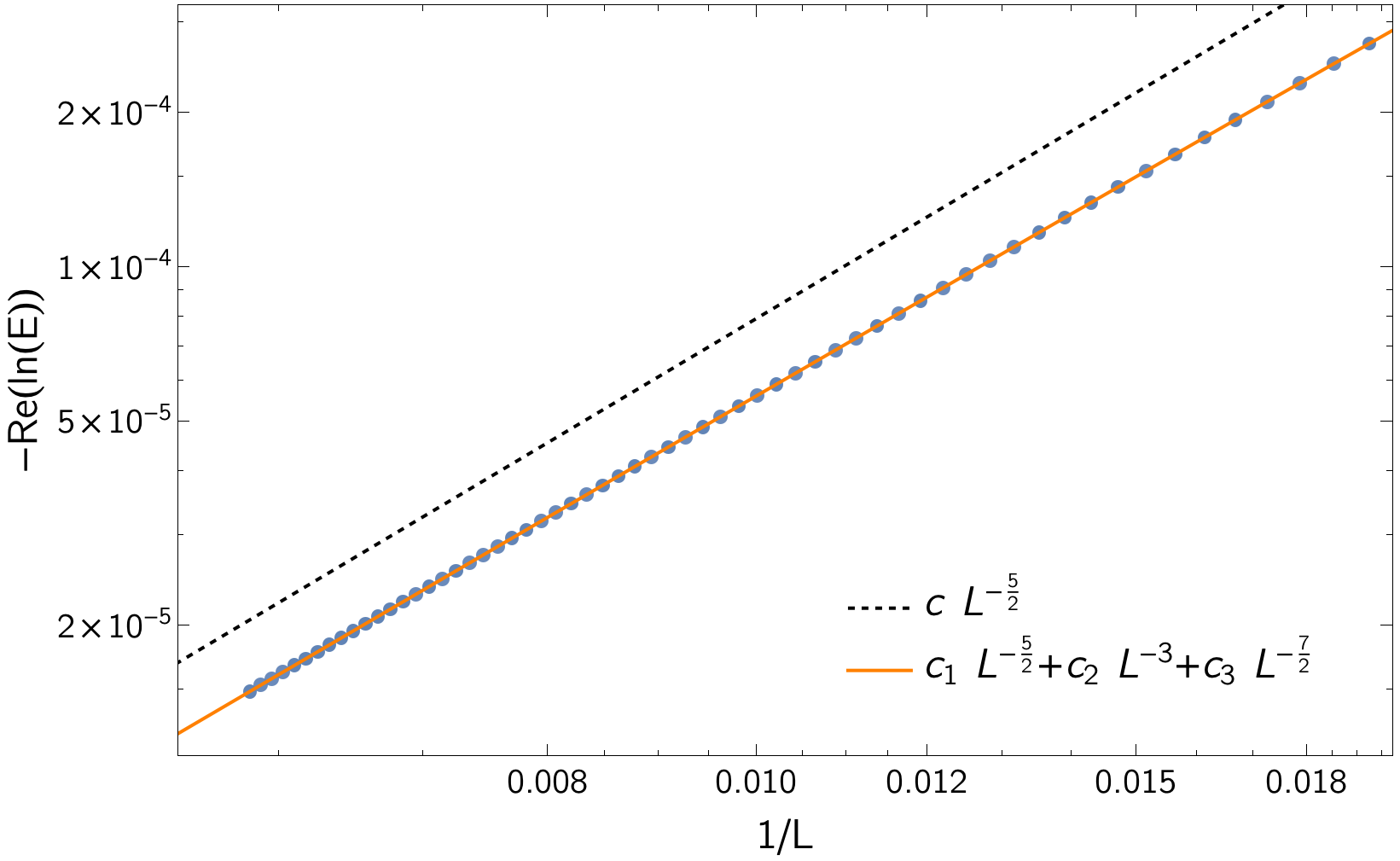}
\caption{Sets of roots $\{u_j|1\leq j\leq L/2\}$ (left panel) and
scaling of the eigenvalue of the transition matrix with system size
(right panel) for the  excited state of  \eq{ints1} for $L=2N$ and
$\alpha=0.1$, with $L=172$ (blue) and $L=92$ (yellow). For large $L$,
the roots approach a non-trivial contour in the complex plane.
The orange line in the right panel is the fit of \eq{fit1_01} to the
functional form of \eq{eofl}.} 
\label{fig:roots_al01}
\end{figure}

%%%%%%%%%%%%%%%%%%%%%%%%%%%%%%%%%%%%%%%%%%%%%%%%%
In \fig{fig:roots_al01}, we show results for State 1 for system
sizes $L\leq 172$. 
As noted above, the imaginary part of the eigenvalue vanishes,
while the $L$-dependence of the real part of $\ln(E)$ is given by \eq{eofl}
with fit parameters
\beq
c_1=6.07924\ ,\quad c_2-1.7243\ ,\quad c_3= -25.554\ .
\label{fit1_01}
\eeq
As shown in \fig{fig:roots_al01}, this provides an excellent
fit to the data.

%%%%%%%%%%%%%%%%%%%%%%%%%%%%%%%%%%%%%%%%%%%%%%%%%
\subsubsection{State 1 at \sfix{$\alpha=0.2$}}
\begin{figure}[ht]
  \centering
  \includegraphics[width=0.45\linewidth]{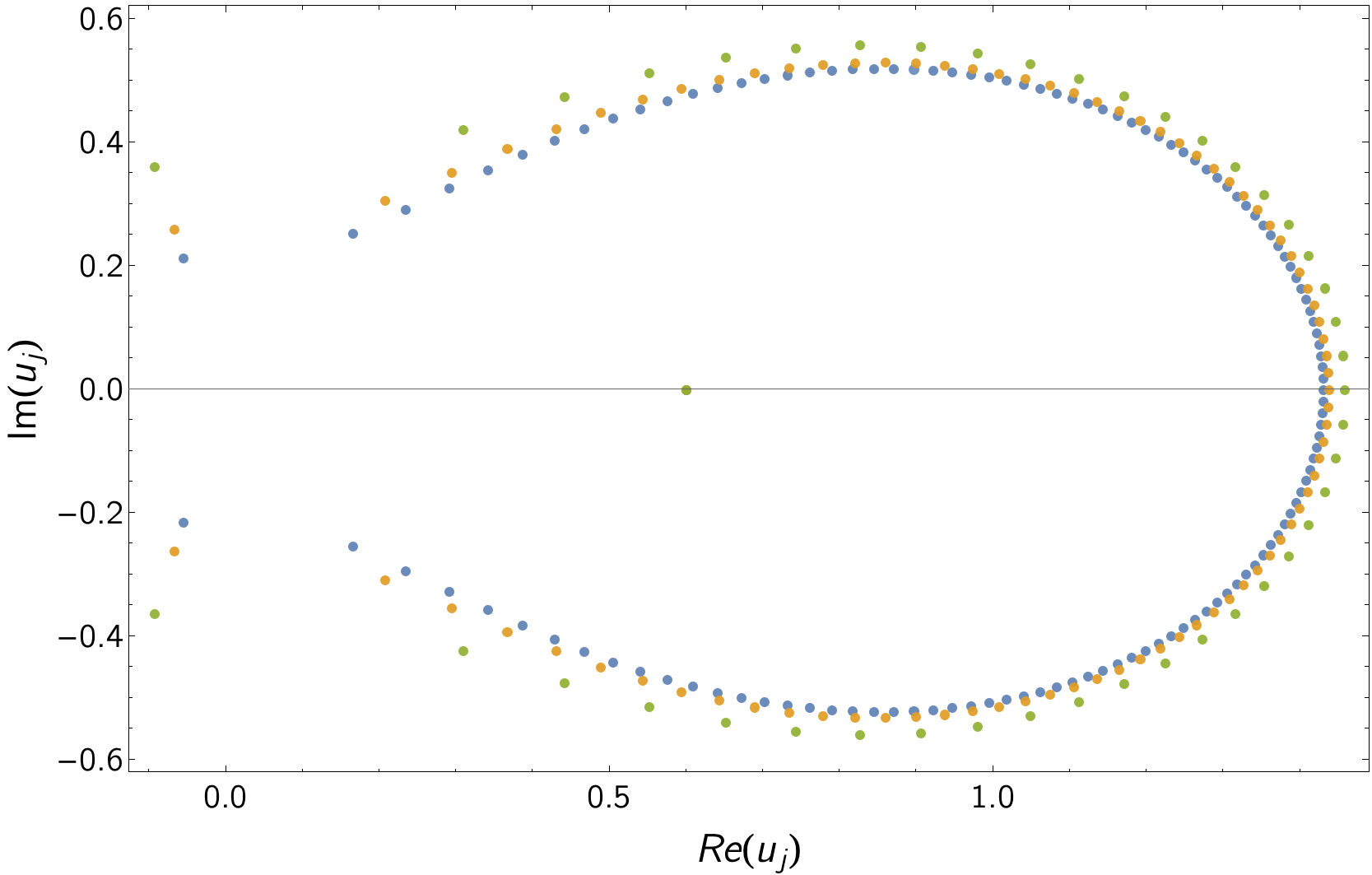}
\qquad
\includegraphics[width=0.45\linewidth]{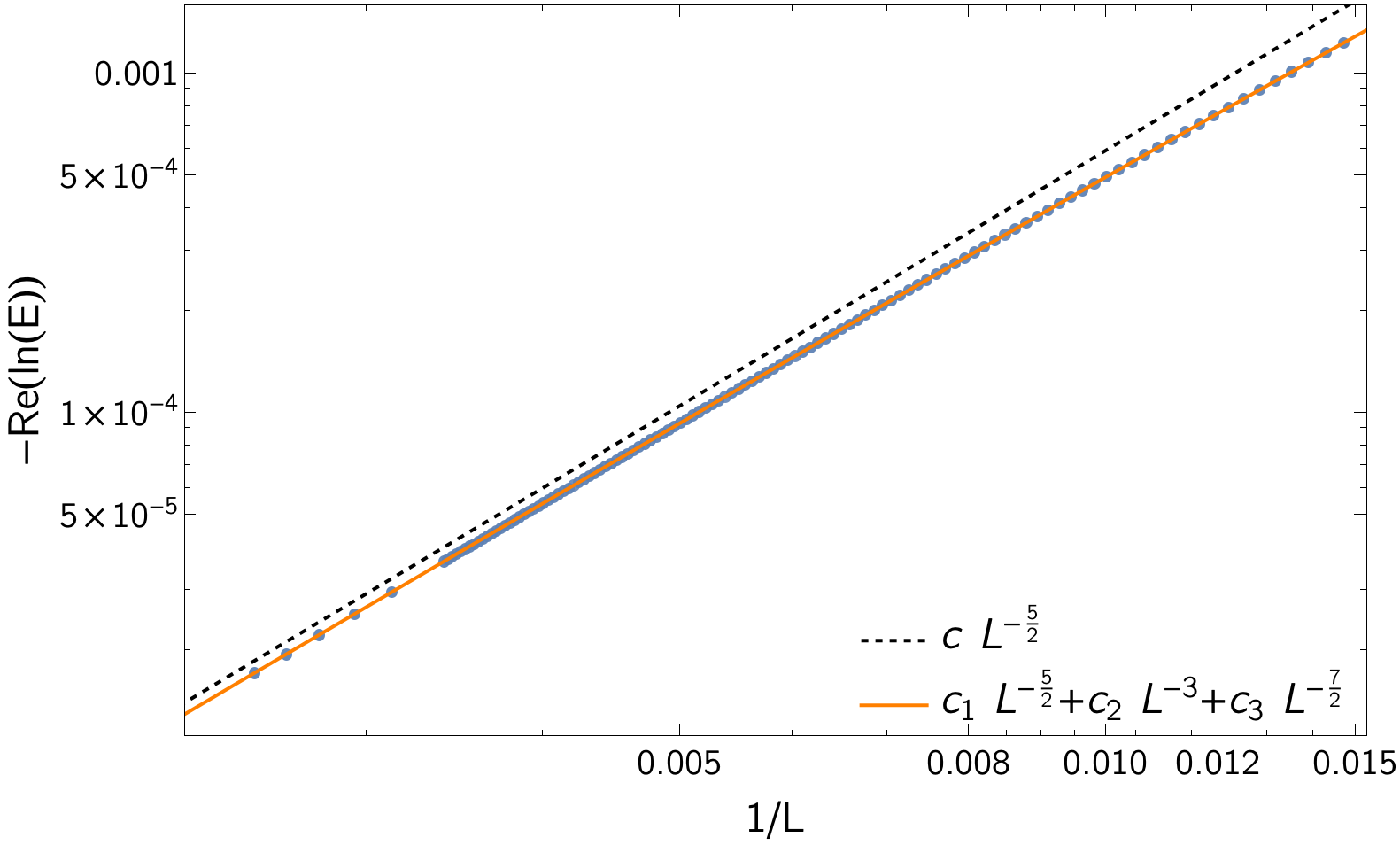}
\caption{Sets of roots $\{u_j|1\leq j\leq L/2\}$ (left panel) and
scaling of the eigenvalue of the transition matrix with system size
(right panel) for the  excited state of \eq{ints1} for $L=2N$ and
$\alpha=0.2$, with $L=248$ (blue), $L=172$ (yellow) and $L=92$
(green). The orange line in the right panel is the fit
of \eq{fit1_02} to the functional form of \eq{eofl}.}
\label{fig:roots_al02}
\end{figure}

%%%%%%%%%%%%%%%%%%%%%%%%%%%%%%%%%%%%%%%%%%%%%%%%%
In \fig{fig:roots_al02}, we show results for State 1 for system
sizes $L\leq 380$. 
As noted above, the imaginary part of the eigenvalue vanishes,
while the $L$-dependence of the real part of $\ln(E)$ is given by \fr{eofl}
with fit parameters
\beq
c_1=58.6837\ ,\quad c_2=-51.4679\ ,\quad c_3=-347.858\ .
\label{fit1_02}
\eeq
As can be seen in \fig{fig:roots_al02}, this provides an excellent
fit to the data. We observe a $L^{5/2}$ scaling of ${\rm
  Re}\big[\ln(E)\big]$ for sufficiently large $L>L_{\rm co}(\alpha)$.
The crossover scale $L^{(1)}_{\rm co}(0.2)$ is significantly larger than the
one for $\alpha=0.1$.

%%%%%%%%%%%%%%%%%%%%%%%%%%%%%%%%%%%%%%%%%%%%%%%%%
\subsubsection{State 1 at \sfix{$\alpha=0.4$}}
\begin{figure}[ht]
  \centering
  \includegraphics[width=0.45\linewidth]{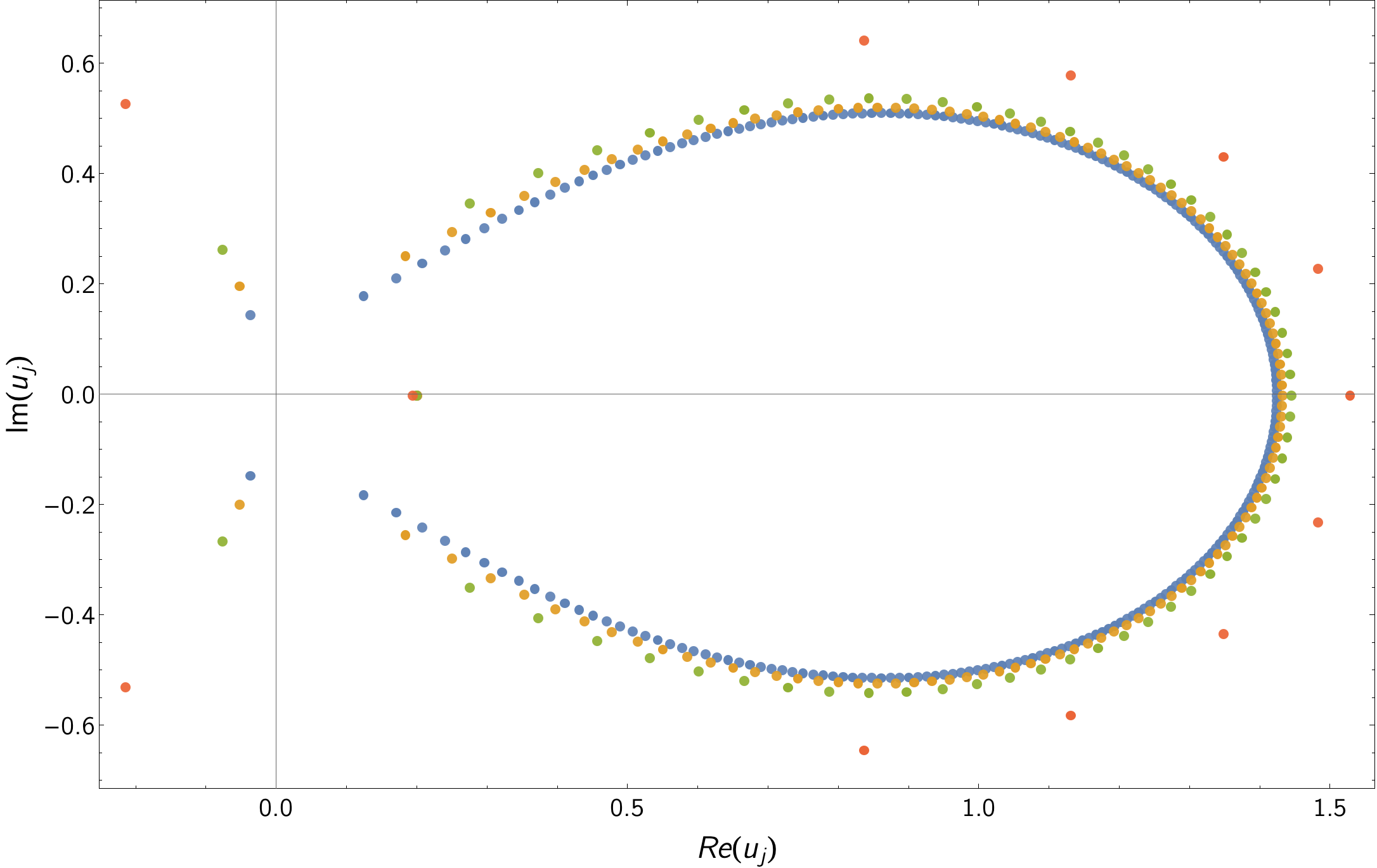}
\qquad
\includegraphics[width=0.45\linewidth]{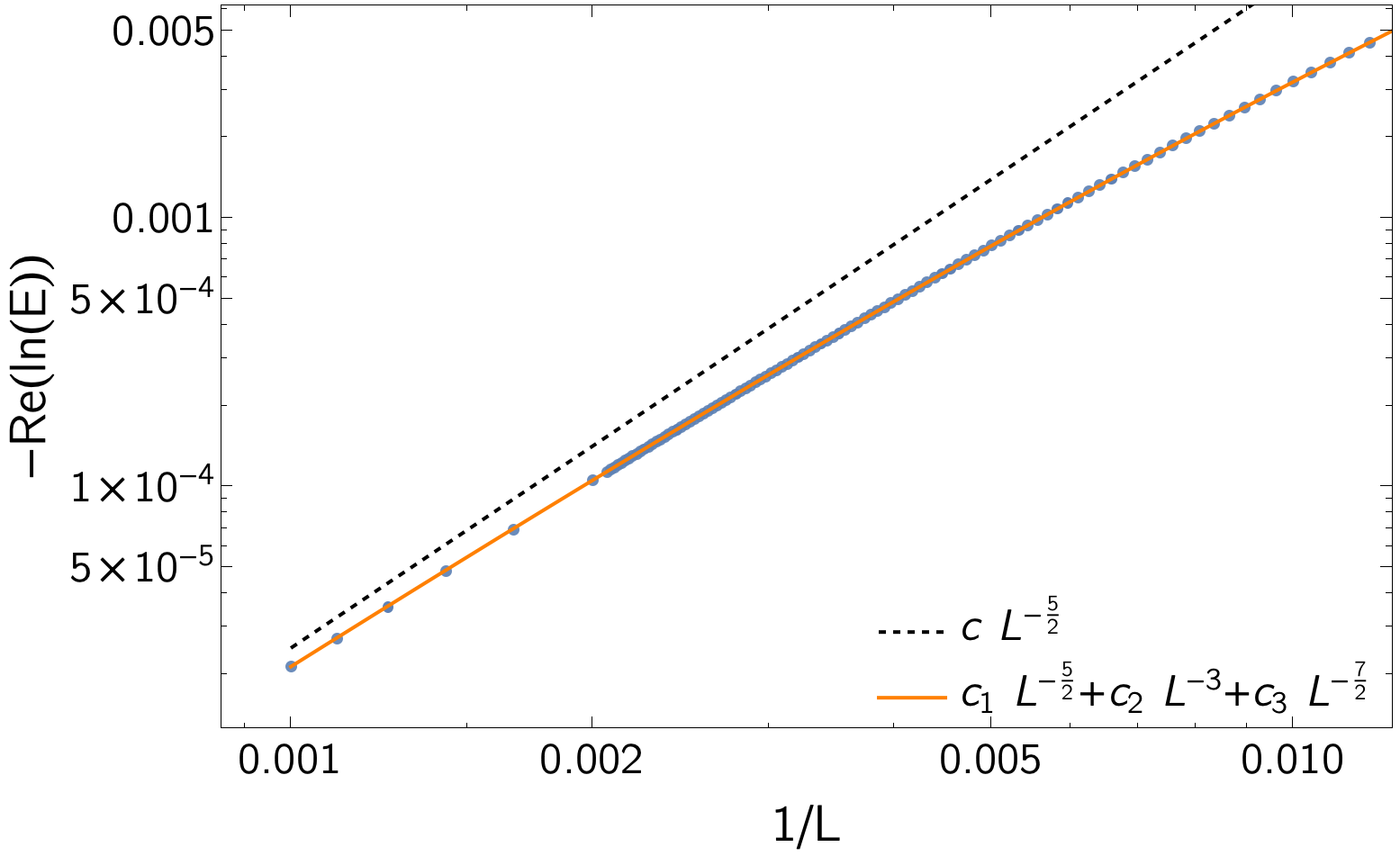}
\caption{Sets of roots $\{u_j|1\leq j\leq L/2\}$ (left panel) and
scaling of the eigenvalue of the transition matrix with system size
(right panel) for the  excited state of  \eq{ints} for $L=2N$ and
$\alpha=0.4$, with $L=484$ (blue), $L=244$ (yellow), $L=124$ (green)
and $L=24$ (red). For large $L$,
the roots approach a non-trivial contour in the complex plane.}
\label{fig:roots_al04}
\end{figure}

%%%%%%%%%%%%%%%%%%%%%%%%%%%%%%%%%%%%%%%%%%%%%%%%%
In \fig{fig:roots_al04}, we show results for State 1 for system
sizes $L\leq 1000$. 
The imaginary part of the eigenvalue vanishes, while the
$L$-dependence of the real part of $\ln(E)$ is given by \fr{eofl}
with fit parameters
\beq
c_1=920.557\ ,\quad c_2=-8286.03\ ,\quad c_3=23384.5\ .
\label{fit1_04}
\eeq
As can be seen in \fig{fig:roots_al04}, this provides an excellent
fit to the data, but the asymptotic $L^{-5/2}$ scaling is approached
only for very large values of $L>L^{(1)}_{\rm co}(0.4)$, where the crossover
scale is much larger than for $\alpha=0.2$ and for $\alpha=0.1$. These
observations are compatible with a crossover scale that diverges as
$\alpha$ approaches the critical value $\alphacrit=1/2$:
\beq
\lim_{\alpha\to \frac{1}{2}}L^{(1)}_{\rm co}(\alpha)= \infty.
\eeq

%%%%%%%%%%%%%%%%%%%%%%%%%%%%%%%%%%%%%%%%%%%%%%%%%
\subsubsection{State 1 at \sfix{$\alpha=0.5$}}
\begin{figure}[ht]
  \centering
  \includegraphics[width=0.45\linewidth]{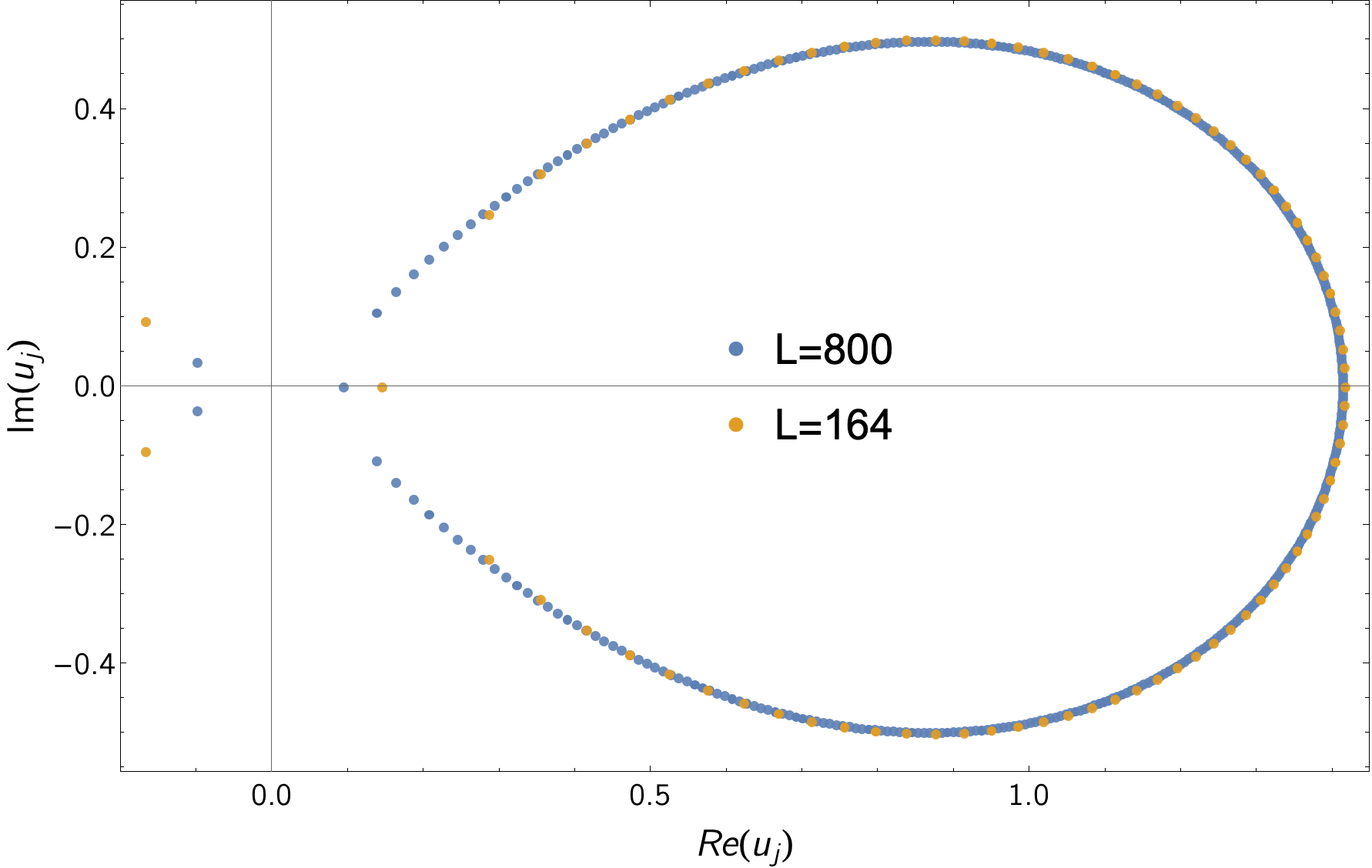}
\qquad
\includegraphics[width=0.45\linewidth]{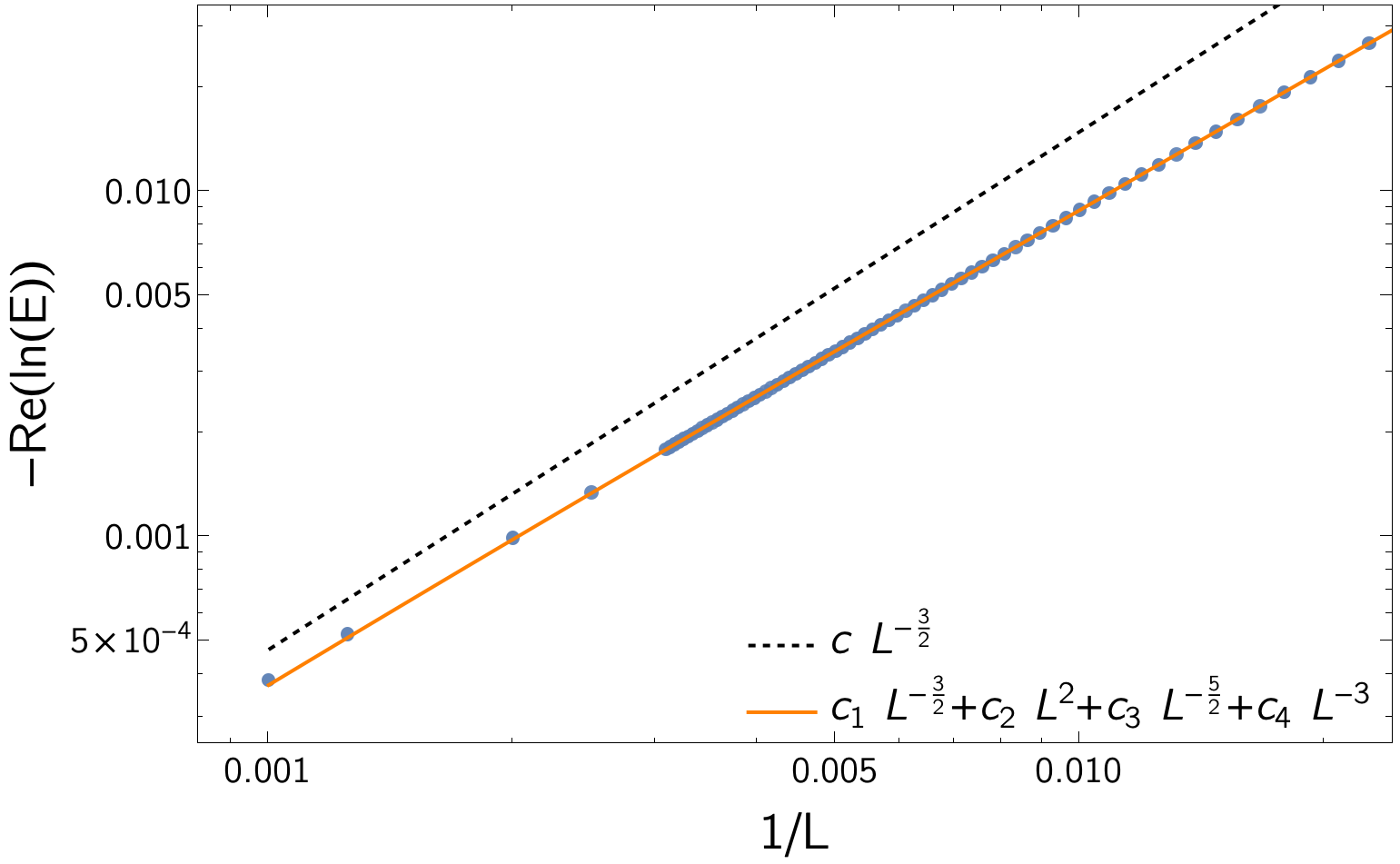}
\caption{Sets of roots $\{u_j|1\leq j\leq L/2\}$ (left panel) and
scaling of the eigenvalue of the transition matrix with system size
(right panel) for the  excited state of \eq{ints1} for $L=2N$ and
$\alpha=0.5$. The orange line in the right panel is the fit of
\eq{fit1_05} to the functional form of \eq{eofl_05a}.}
\label{fig:roots_al05_state1}
\end{figure}

%%%%%%%%%%%%%%%%%%%%%%%%%%%%%%%%%%%%%%%%%%%%%%%%%
In Fig.~\ref{fig:roots_al05_state1}, we show results for State 1 for system
sizes $L\leq 1000$. 
The imaginary part of the eigenvalue is zero, while the $L$-dependence
of the real part of $\ln(E)$ is fitted to the functional form
\beq
E(L)=c_1L^{-3/2}+c_2L^{-2}+c_3L^{-5/2}+c_4L^{-3}\ ,
\label{eofl_05a}
\eeq
where
\beq
c_1=14.3328\ ,\quad c_2=-93.2641\ ,\quad c_3=493.579\ ,\quad c_4=-1052.21\ .
\label{fit1_05}
\eeq
As can be seen in \fig{fig:roots_al05_state1}, this provides an excellent
fit to the data. Our numerical results are compatible with an
asymptotic $L^{-3/2}$ scaling.

%%%%%%%%%%%%%%%%%%%%%%%%%%%%%%%%%%%%%%%%%%%%%%%%%
\subsubsection{State 3 at \sfix{$\alpha=0.5$}}
\begin{figure}[ht]
  \centering
  \includegraphics[width=0.45\linewidth]{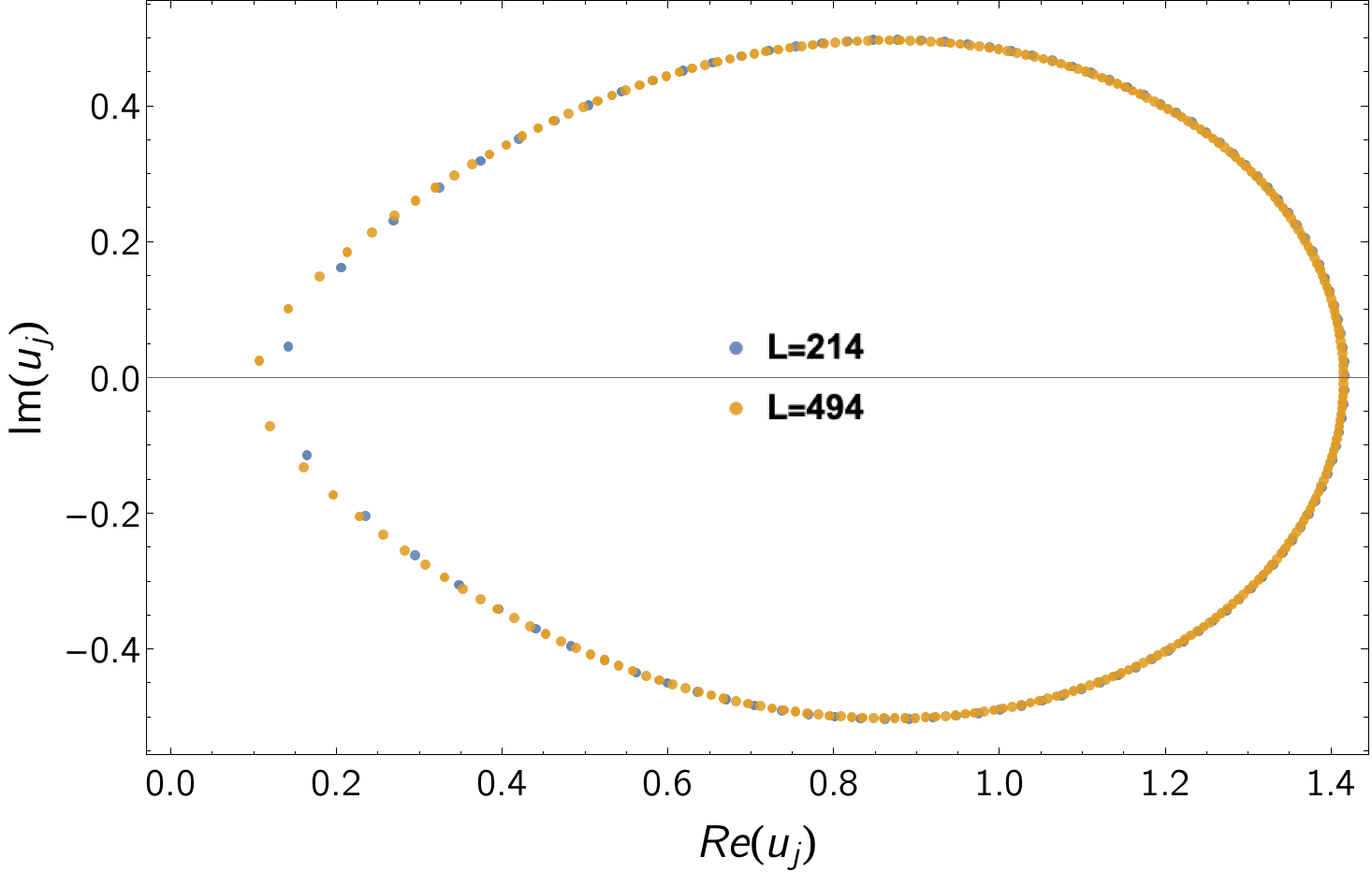}
\qquad
\includegraphics[width=0.45\linewidth]{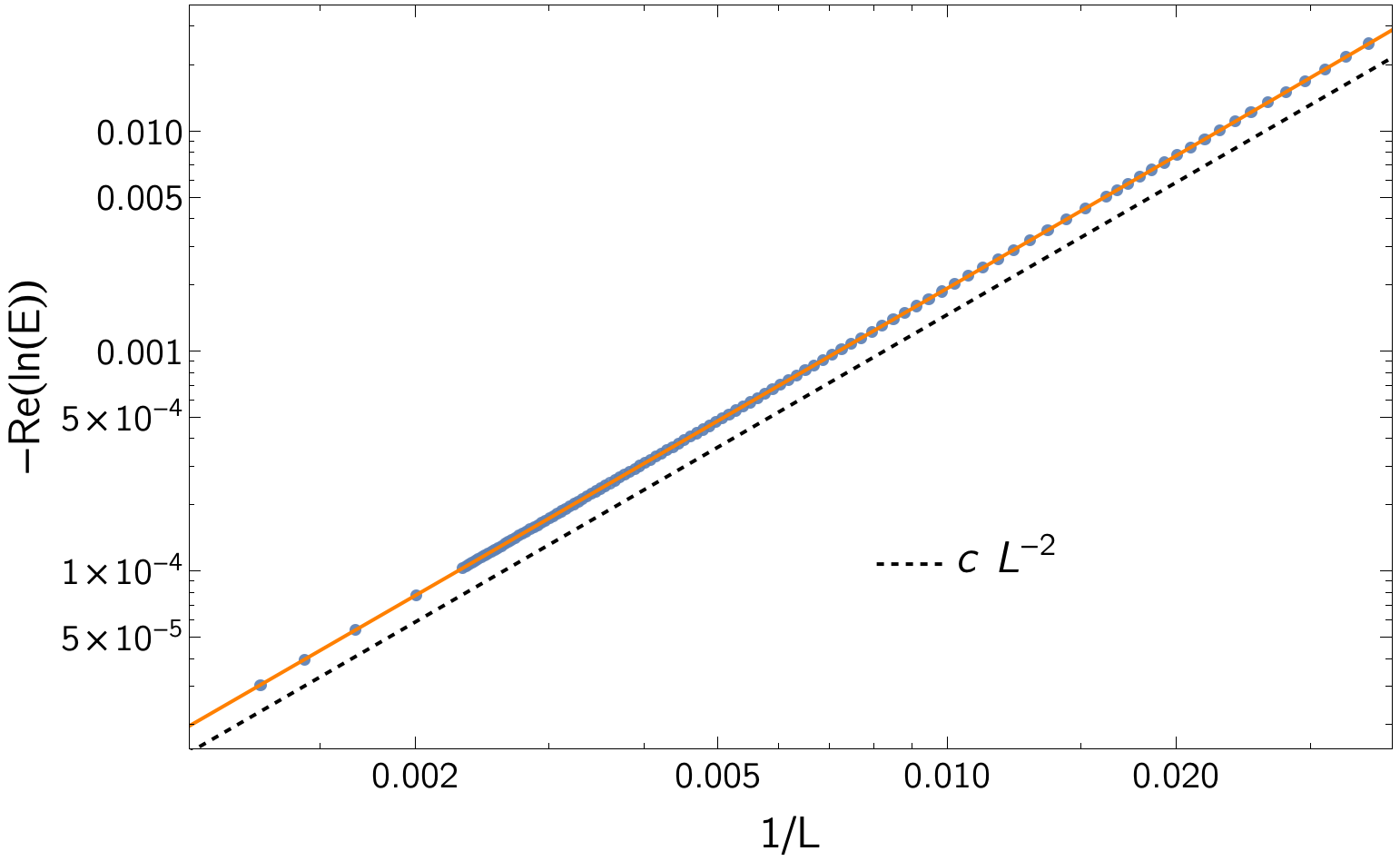}
\caption{Sets of roots $\{u_j|1\leq j\leq L/2\}$ (left panel) and
scaling of the eigenvalue of the transition matrix with system size
(right panel) for the  excited state of \eq{ints3} for $L=2N$ and
$\alpha=0.5$. The orange line in the right panel is the fit
of \eq{fit3} to the functional form of \eq{eofl_05}.}
\label{fig:roots_al05_state3}
\end{figure}

%%%%%%%%%%%%%%%%%%%%%%%%%%%%%%%%%%%%%%%%%%%%%%%%%
In Fig.~\ref{fig:roots_al05_state3} we show results for State 3 for system
sizes $L\leq 800$. 
The imaginary part of the eigenvalue vanishes, while the
$L$-dependence of the real part of $\ln(E)$ is fitted to the
functional form
\beq
E(L)=c_1L^{-2}+c_2L^{-5/2}+c_3L^{-3}\ ,
\label{eofl_05}
\eeq
where
\beq
c_1=19.9061\ ,\quad c_2= -3.49057\ ,\quad c_3=20.97\ .
\label{fit3}
\eeq
As can be seen in \fig{fig:roots_al05_state3}, this provides an excellent
fit to the data.

%%%%%%%%%%%%%%%%%%%%%%%%%%%%%%%%%%%%%%%%%%%%%%%%%
\subsubsection{State 2 at \sfix{$\alpha=0.7$}}
\begin{figure}[ht]
  \centering
  \includegraphics[width=0.45\linewidth]{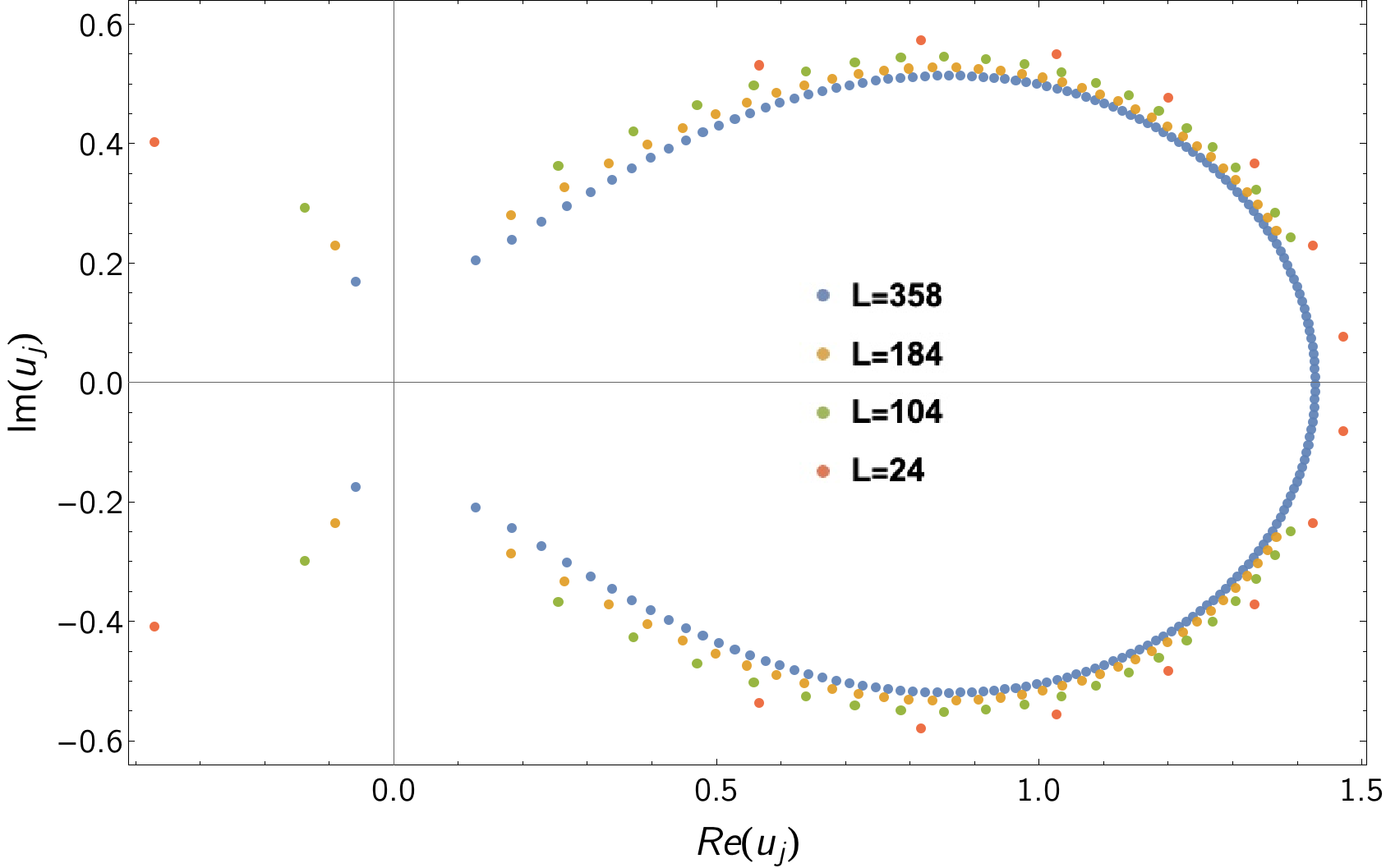}
\qquad
\includegraphics[width=0.45\linewidth]{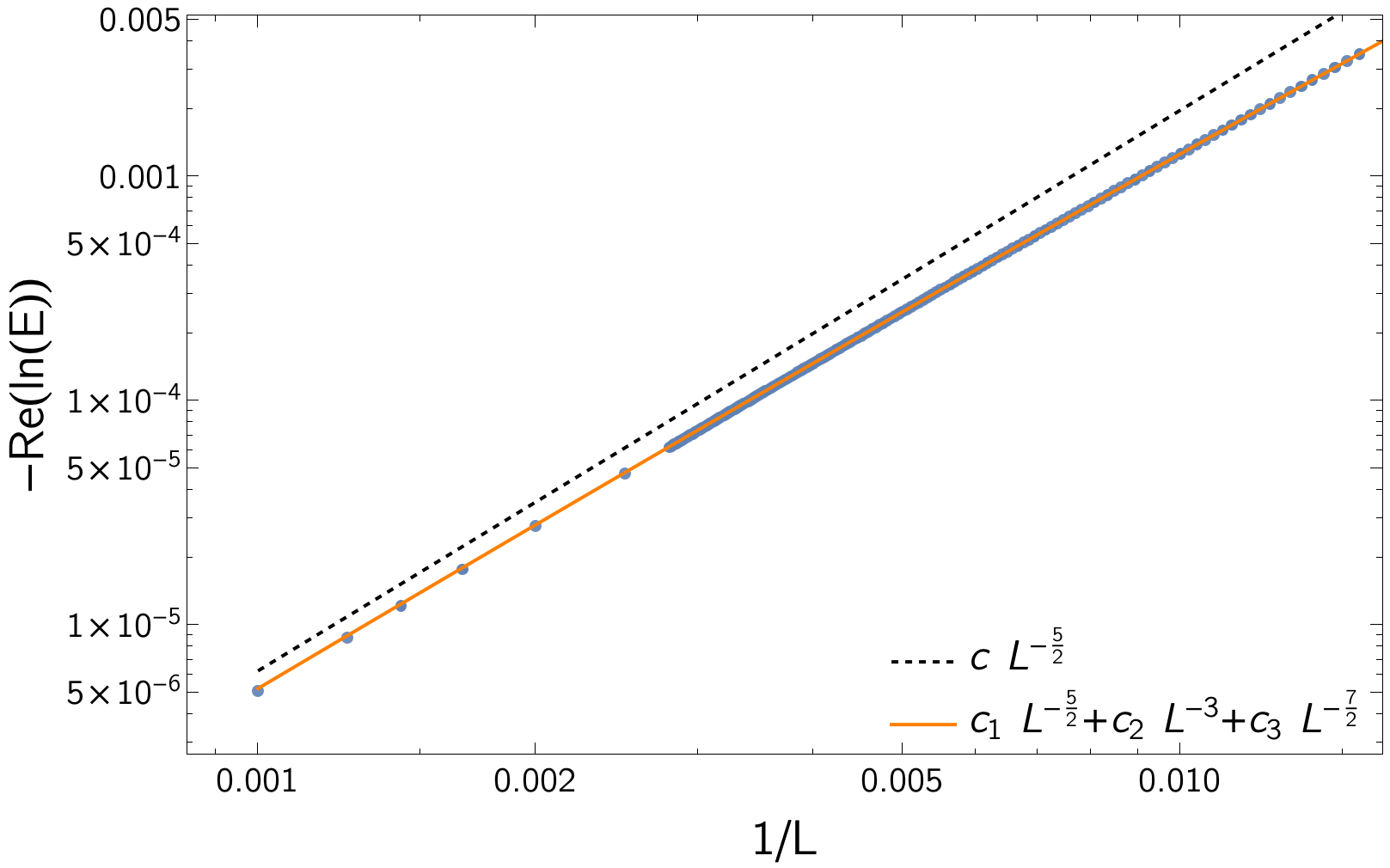}
\caption{Sets of roots $\{u_j|1\leq j\leq L/2\}$ (left panel) and
scaling of the eigenvalue of the transition matrix with system size
(right panel) for the  excited state of  \eq{ints} for $L=2N$ and
$\alpha=0.7$. The orange line in the right panel is the fit
of \eq{fit1_07} to the functional form of
\eq{eofl}.}
\label{fig:roots_al07}
\end{figure}

%%%%%%%%%%%%%%%%%%%%%%%%%%%%%%%%%%%%%%%%%%%%%%%%%
In Fig.~\ref{fig:roots_al07} we show results for State 2 for system
sizes $L\leq 1000$. 
The imaginary part of the eigenvalue vanishes, while the
$L$-dependence of the real part of $\ln(E)$ is given by \eq{eofl}
with fit parameters
\beq
c_1=189.245\ ,\quad c_2=-750.725\ ,\quad c_3=1353.99\ .
\label{fit1_07}
\eeq
As can be seen in \fig{fig:roots_al07}, this provides an excellent
fit to the data, but the asymptotic $L^{-5/2}$ scaling is approached
only for large values of $L>L^{(2)}_{\rm co}(0.7)$, where the crossover
scale is much larger than for $\alpha=0.8$.

%%%%%%%%%%%%%%%%%%%%%%%%%%%%%%%%%%%%%%%%%%%%%%%%%
\subsubsection{\sfix{$\alpha=0.8$}}
\begin{figure}[ht]
  \centering
  \includegraphics[width=0.45\linewidth]{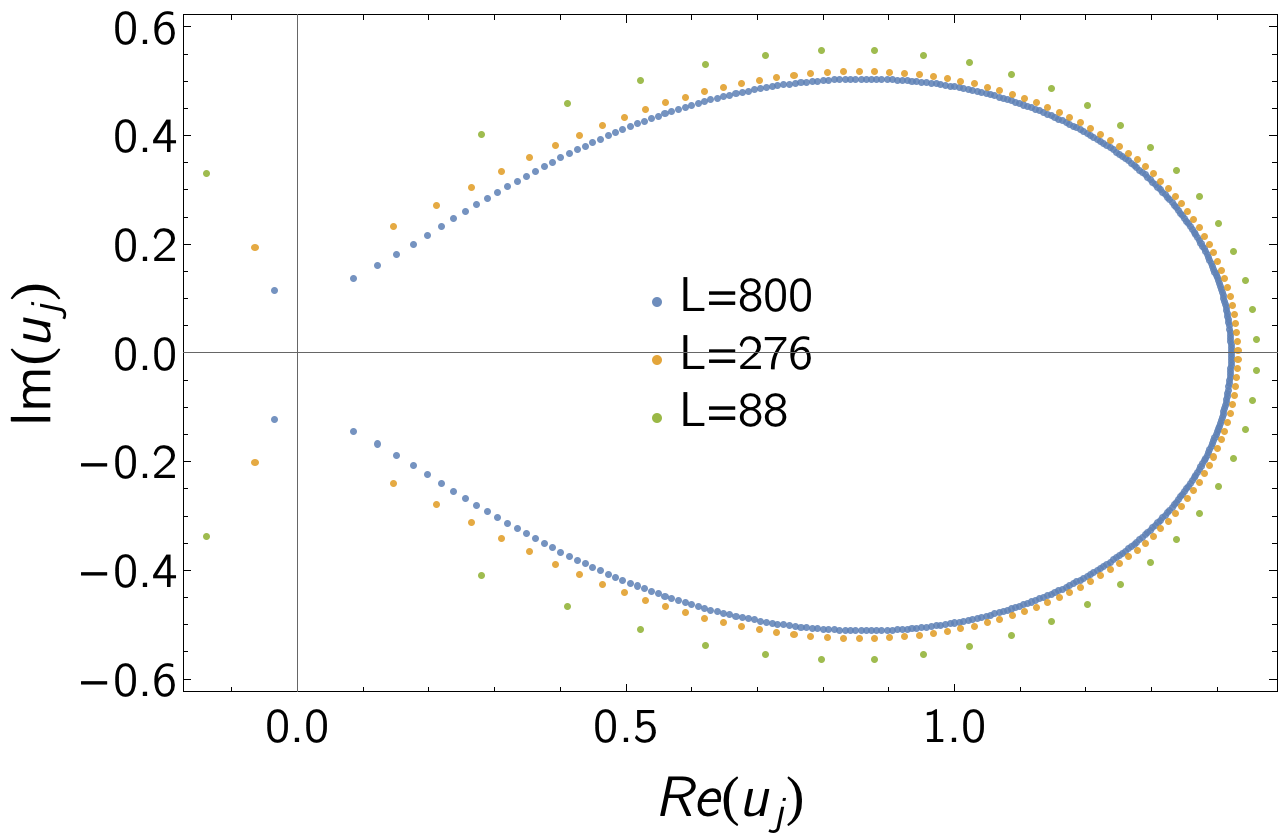}
\qquad
\includegraphics[width=0.45\linewidth]{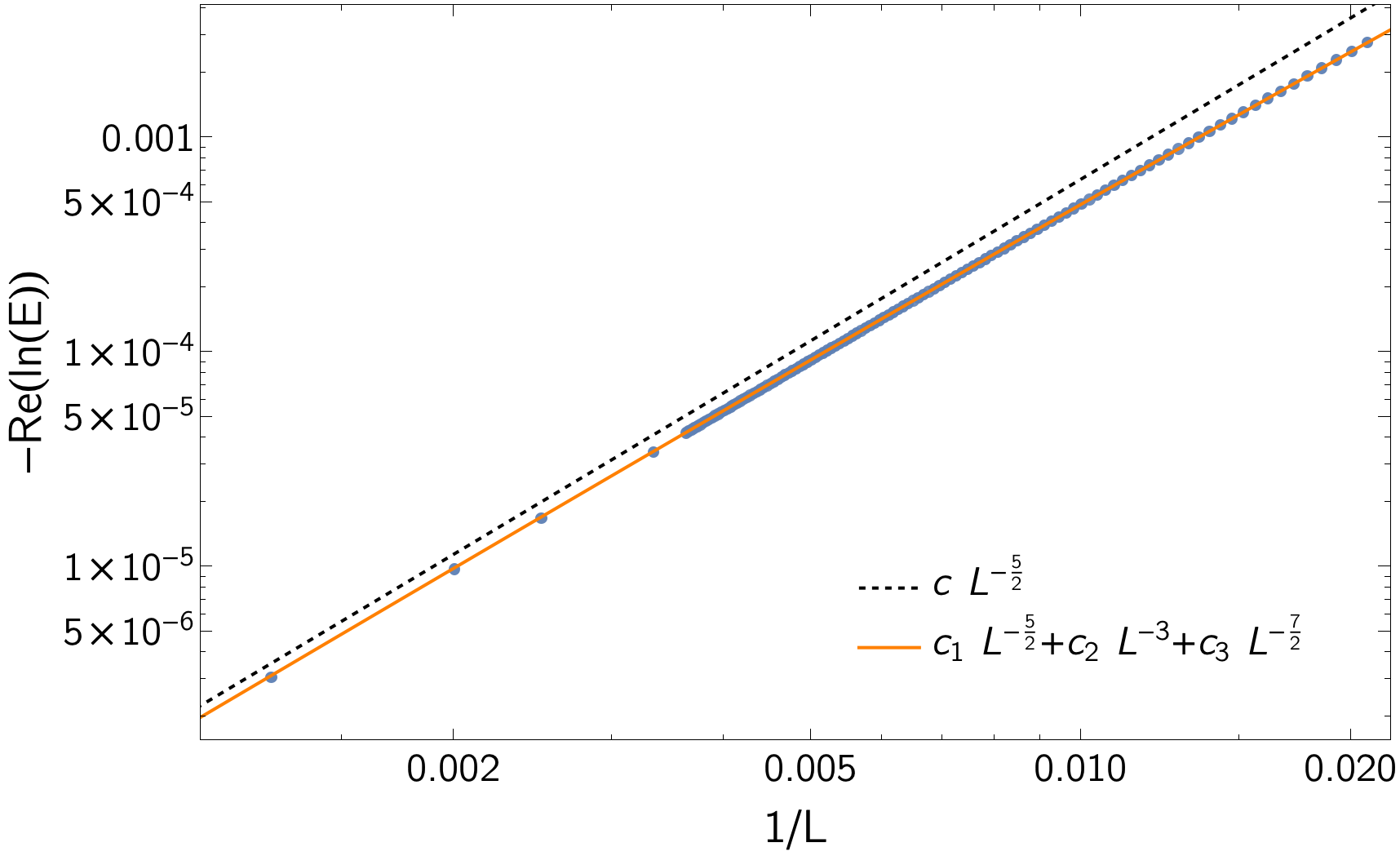}
\caption{Sets of roots $\{u_j|1\leq j\leq L/2\}$ (left panel) and
scaling of the eigenvalue of the transition matrix with system size
(right panel) for  the  excited state of  \eq{ints} for $L=2N$ and
$\alpha=0.8$. The roots are shown for $L=800$ (blue),  $L=276$
(yellow) and $L=88$ (green). }
\label{fig:roots_al08}
\end{figure}

%%%%%%%%%%%%%%%%%%%%%%%%%%%%%%%%%%%%%%%%%%%%%%%%%
In \fig{fig:roots_al08}, we show results for State 2 for system
sizes $L\leq 800$. 
The imaginary part of the eigenvalue vanishes, while the
$L$-dependence of the real part of $\ln(E)$ is given by \eq{eofl}
with fit parameters
\beq
c_1=61.2629\ ,\quad c_2= -122.315\ ,\quad c_3=57.1236\ .
\label{fit1_08}
\eeq
As can be seen in \fig{fig:roots_al08}, this provides an excellent
fit to the data.

%%%%%%%%%%%%%%%%%%%%%%%%%%%%%%%%%%%%%%%%%%%%%%%%%
\subsubsection{Excitation gap}
%%%%%%%%%%%%%%%%%%%%%%%%%%%%%%%%%%%%%%%%%%%%%%%%%
The above analysis provides us with bounds for the
spectral gap 
\beq
\Delta={\rm Re}\ln[E^*]\ ,
\eeq
where $E^* \neq 1$ is the eigenvalue of the transition
matrix with magnitude closest to one.
We conclude that,  for large $L$, we must have
\beq
\Delta\leq \begin{cases}
  {\rm const}\ L^{-5/2} & \text{ if } \alpha\neq\frac{1}{2}\ ,\\
  {\rm const}\ L^{-2} & \text{ if } \alpha=\frac{1}{2}\ .\\
  \end{cases}
\eeq
%This was discussed in detail in \REF{essler2024lifted}.

%%%%%%%%%%%%%%%%%%%%%%%%%%%%%%%%%%%%%%%%%%%%%%%%%%%%%%%%%%%%%%%%%%%%%%%%%%%%%%%%%%%
\section{Resolving the discrepancy between Bethe Ansatz and MC simulations}
\label{sec:resolution}
%%%%%%%%%%%%%%%%%%%%%%%%%%%%%%%%%%%%%%%%%%%%%%%%%%%%%%%%%%%%%%%%%%%%%%%%%%%%%%%%%%%
The bound for the spectral gap of the transition matrix obtained from
the Bethe ansatz solution
\beq
\Delta\glb\alpha=\frac{1}{2} \grb\leq {\rm const}\ L^{-2}\ ,
\eeq
differs from the autocorrelation time of the structure factor and
similar observables, which, according to our MC simulations, behave as
\beq
\tau_{\rm IAC} \sim L^{3/2}\ .
\eeq
A possible explanation is that the eigenvector(s) that corresponds to
eigenvalues such that ${\rm Re}\ln(E)\propto L^{-2}$ have negligible  overlap
with the large-scale density modes tracked by the structure factor, and there
are too few of them.
%%%%%%%%%%%%%%%%%%%%%%%%%%%%%%%%%%%
\subsection{Spectral representation of linear-response functions}
%%%%%%%%%%%%%%%%%%%%%%%%%%%%%%%%%%%
In order to proceed, it is useful to view the configurations
$\{\vec{j},n\}$ as orthonormal basis vectors in a linear vector
space 
\beq
|\vec{j};n\rangle\ ,\qquad
\langle\vec{k};m|\vec{j};n\rangle=\delta_{n,m}\delta_{\vec{k},\vec{j}}\ .
\label{basisstates}
\eeq
We then define the uniform state by
\beq
|u\rangle=\sum_{\vec{j},n} |\vec{j};n\rangle\ .
\eeq
The normalized steady state for $L=2N$ is 
\beq
|P_{\rm SS}\rangle=\frac{1}{{\cal N}}|u\rangle\ ,\quad {\cal N}=\frac{L}{2}\begin{pmatrix}
L\\ L/2
\end{pmatrix}.
\eeq
The master equation then reads
\beq
\frac{d}{dt}|P(t)\rangle= \hat{T}|P(t)\rangle.
\eeq
The observables of interest act on states as
\beq
\hat{\cal A}|\vec{j};n\rangle=\nu_{\cal A}(\vec{j},n)
|\vec{j};n\rangle\ .
%\nu_{\cal A}(\vec{j},n)={\cal O}(L^0)\ .
\eeq
Dynamical susceptibilities in equilibrium can then be written in the form 
\beq
\chi_{{\cal AB}}(t)=\langle u|\hat{\cal A}\hat{T}^t\hat{\cal B}|P_{\rm SS}\rangle\ .
\eeq
A particular example considered in Ref.~\cite{essler2024lifted} and
discussed above is
\beq
{\cal A}={\cal B}=\hat{S}(-Q)\hat{S}(Q)\ ,\quad
\hat{S}(Q)|\vec{j};n\rangle=\frac{1}{\sqrt{L}}\sum_{n=1}^Ne^{iQj_n}|\vec{j};n\rangle\ .
\label{sq}
\eeq
It is useful to express these in a spectral representation in terms of
eigenstates of the transition matrix $\hat{T}$. Assuming that
$\hat{T}$ is diagonalizable by a similarity transformation (which we
observe to hold for small $L$), we can take 
\begin{align}
\hat{T}|R_n\rangle=E_n|R_n\rangle\ ,\quad
  \langle L_n|T=E_n\langle L_n|\ ,\quad \langle
  L_n|R_m\rangle=\delta_{n,m}\ ,\quad
  \mathbb{1}=\sum_n|R_n\rangle\langle L_n|\ .
\label{LREV}
\end{align}
This leads to the following spectral representation of the dynamical susceptibility
\begin{align}
  \chi_{{\cal AB}}(t)=\sum_n \langle u|\hat{\cal A}|R_n\rangle
  \langle L_n|\hat{\cal B}|P_{\rm SS}\rangle\ 
e^{t\ln(E_n)}\ .
\end{align}
In practice we will normalize the right eigenstates
\beq
\langle R_n|R_n\rangle=1\ ,
\eeq
and the normalization of the $\langle L_n|$ then follows from \fr{LREV}.
Our analysis of the Bethe ansatz equations suggests that the spectral
gap scales as a power-law in $L$ 
\beq
\Delta\propto L^{-\alpha}\ .
\eeq
Hence, in order to measure the correlation time $\tau_{\rm corr}\sim
\Delta^{-1}$ in Monte Carlo simulation we need to consider times such
that
\beq
t> L^\alpha\ .
\eeq
As we are dealing with an interacting many-particle system, typical matrix
elements are expected to scale as
\beq
\langle u|\hat{\cal A}|R_n\rangle
\langle L_n|\hat{\cal B}|P_{\rm SS}\rangle=
{\cal O}\big(e^{-\gamma L}\big)\ .
\eeq
As exponentially small (in $L$) effects are not accessible in
Monte Carlo simulations on the large systems we are interested in, the
$L$-dependence of the correlation time $\tau_{\rm corr}$ can be
measured only if one of the following conditions is met:
\begin{enumerate}
\item{} There exists an exponentially large number of eigenstates
  $|R_n\rangle$ whose eigenvalues $E_n$ have the same scaling as the spectral gap
\beq
{\rm Re}\big(\ln(E_n)\big)\propto L^{-\alpha}\ .
\eeq
Summing over them in the spectral representation then gives a ${\cal
  O}(L^0)$ contribution to the susceptibility that at behaves as
\begin{align}
  \chi_{{\cal AB}}(t)\sim c_1 e^{-c_2 t/L^{\alpha}}\ ,\quad
    t>L^\alpha\ .
\label{scaling}
\end{align}
\item{} The number of eigenstates that have the same scaling as the
spectral gap scales as a power of the system size, but they give rise
to atypically large matrix elements
\beq
\langle u|\hat{\cal A}|R_n\rangle
\langle L_n|\hat{\cal B}|P_{\rm SS}\rangle=
{\cal O}\big(L^{-\beta}\big)\ .
\eeq
In this case summing over these eigenstates again gives a contribution
to the susceptibility of the form of \eq{scaling}, which dominates the
late-time behaviour.
\end{enumerate}
We now will present some results that support neither scenario
in the \LT at $\alpha=\alphacrit$. This suggests that the
relaxation time (the inverse gap) is not visible in the decay
of dynamical susceptibilities and related observables extracted from
Monte Carlo simulations.

%%%%%%%%%%%%%%%%%%%%%%%%%%%%%%%%%%%%%%%%%%%%%%%%%%%%%%%%%%%%%%%%
\subsection{Translational invariance and momentum eigenbasis}
\label{sec:mtmbasis}
%%%%%%%%%%%%%%%%%%%%%%%%%%%%%%%%%%%%%%%%%%%%%%%%%%%%%%%%%%%%%%%%
The translation operator $\hat{\tau}$ acts on basis states as
\beq
\hat{\tau}|\vec{j};n\rangle=\begin{cases}
|j_1+1,\dots,j_N+1;n\rangle &\text{if } j_N<L\ ,\\
|1,j_1+1,\dots,j_{N-1}+1;n+1\rangle &\text{if } j_N=L\ .
\end{cases}
\eeq
Here, the pointer variable $n+1$ is to be understood mod $L/2$. As we
are working on an $L$-site lattice with periodic boundary conditions, 
we have $\hat{\tau}^L=\mathbb{1}$.
A basis of momentum eigenstates is then given by
\begin{align}
|q;\vec{j}_0\rangle&=\frac{1}{\sqrt{L}}\sum_{k=0}^{L-1}e^{iqk}\hat{\tau}^k|\vec{j}_0;1\rangle\ ,\quad
q=\frac{2\pi m}{L}\ ,\ -\frac{L}{2}< m\leq \frac{L}{2}\ ,\\
\hat{\tau}|q;\vec{j}_0\rangle&=e^{-iq}|q;\vec{j}_0\rangle\ ,
\label{mtmEVs}
\end{align}
where $\{\vec{j}_0;1\}$ is the set of configurations with the pointer
particle located on site $1$. This basis is useful because the
Bethe ansatz states are by construction momentum eigenstates and we
can gain insight into their structure by looking at them in the
$\{|q;\vec{j}_0\rangle\}$ basis.

%%%%%%%%%%%%%%%%%%%%%%%%%%%%%%%%%%%%%%%%%%%%%%%%%%%%%%%%%%%%%%%%
\subsubsection{Action of observables on momentum eigenstates}
\label{ssec:obstrans}
%%%%%%%%%%%%%%%%%%%%%%%%%%%%%%%%%%%%%%%%%%%%%%%%%%%%%%%%%%%%%%%%
The dynamical structure factor \fr{sq} is translationally invariant
\beq
\hat{\tau}\hat{S}(-Q)\hat{S}(Q)|q;\vec{j}_0\rangle=
e^{-iq}\hat{S}(-Q)\hat{S}(Q)|q;\vec{j}_0\rangle\ ,
\eeq
which implies that
\beq
\langle u|\hat{S}(-Q)\hat{S}(Q)|q;\vec{j}_0\rangle\propto \delta_{q,0}\ .
\eeq
Hence the eigenstate exhibiting $L^{-2}$ scaling (State 3) considered
above does not contribute in the spectral representation of
$\chi_{\cal AA}$ with ${\cal A}=\hat{S}(-Q)\hat{S}(Q)$.
However, the observable $\hat{S}(Q)$ (as well as other observables we
have considered in our Monte Carlo simulations) does have
non-vanishing matrix elements between $\langle u|$ and finite-momentum
states 
\beq
\langle u|\hat{S}(Q)|q;\vec{j}_0\rangle\propto \delta_{Q,-q}\ .
\eeq
%%%%%%%%%%%%%%%%%%%%%%%%%%%%%%%%%%%%%%%%%%%%%%%%%
\subsection{Structure of State 3 for small $L=2N$}
\begin{figure}[ht]
\centering
\includegraphics[width=0.45\linewidth]{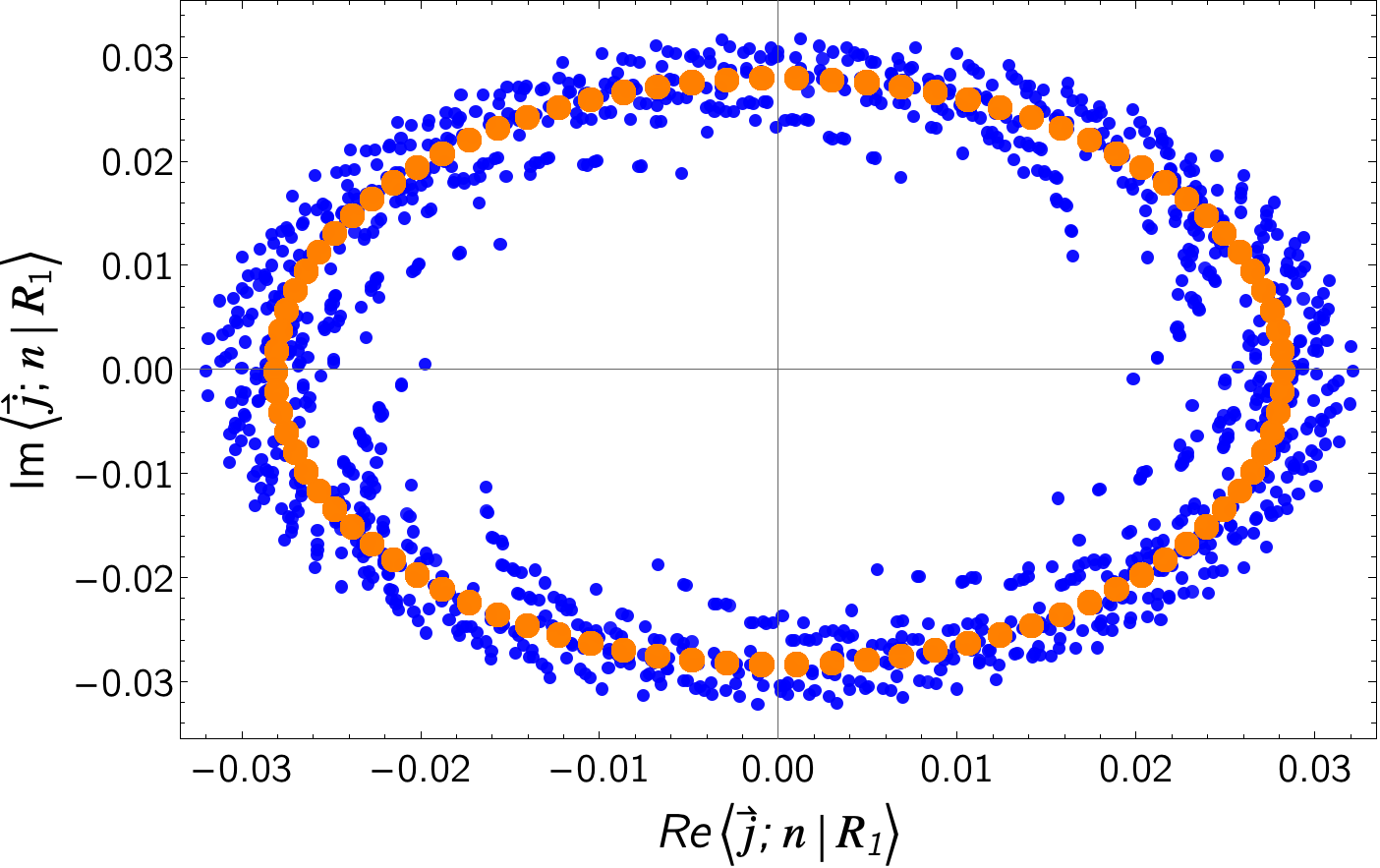}\
\includegraphics[width=0.45\linewidth]{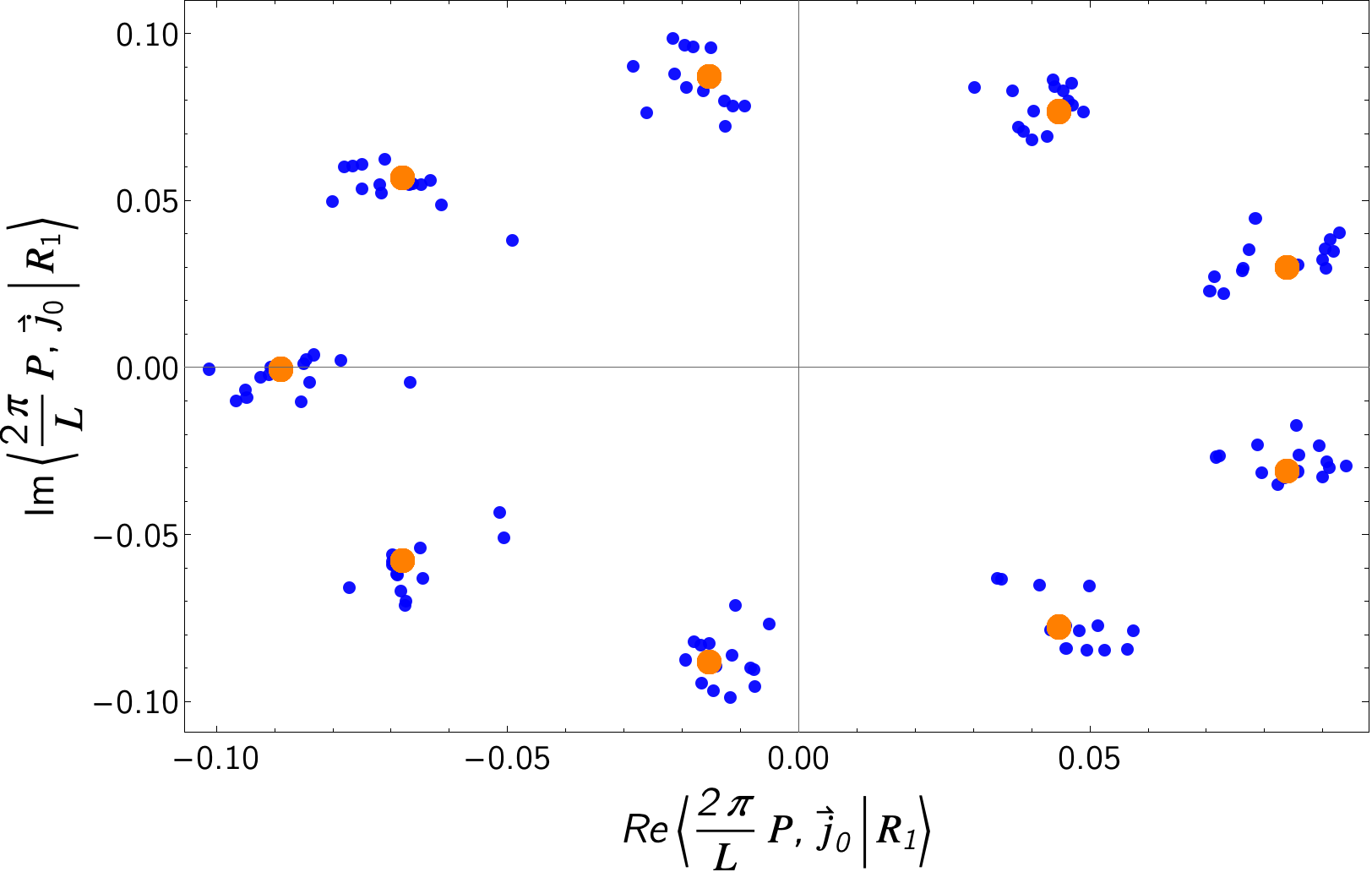}
\caption{Plot of the real and imaginary parts of the amplitudes
$\langle\vec{j};m|R_1\rangle$ (left panel, blue crosses) and
$\langle\frac{2\pi}{L};\vec{j}_0|R_1\rangle$ (right panel), where
$|R_1\rangle$ is the eigenstate \fr{rootsN5} ($L=2N=10$). The orange
dots are, respectively, the amplitudes
$\langle\vec{j};m|F_{1,\psi}\rangle$ and
$\langle\frac{2\pi}{L};\vec{j}_0|F_{1,\psi}\rangle$ for the
variational state  defined in \eq{FVector} with
$\psi=\pi/4$.}
\label{fig:N5}
\end{figure}

\begin{figure}[ht]
\centering
\includegraphics[width=0.3\linewidth]{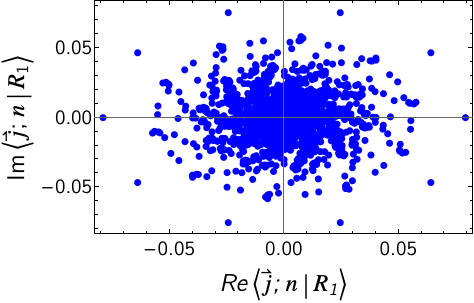}
\includegraphics[width=0.3\linewidth]{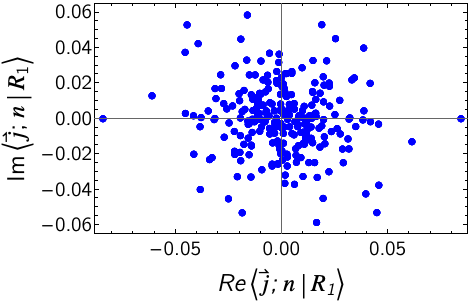}
\includegraphics[width=0.3\linewidth]{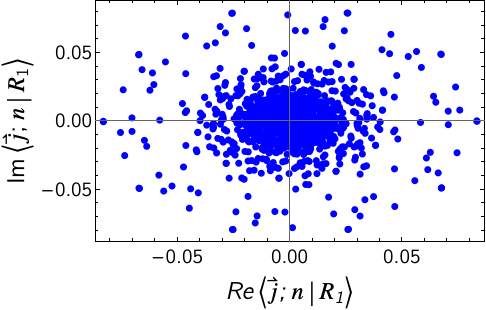}
\\
\includegraphics[width=0.3\linewidth]{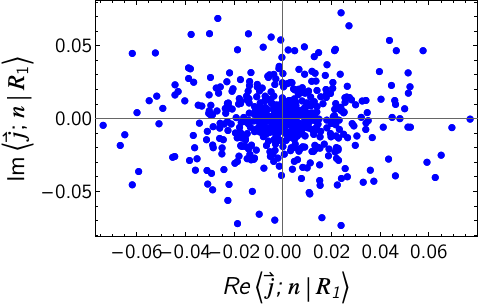}
\includegraphics[width=0.3\linewidth]{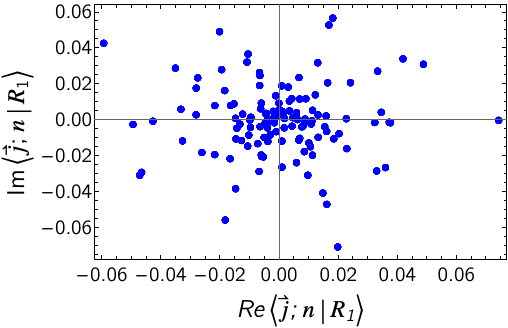}
\includegraphics[width=0.3\linewidth]{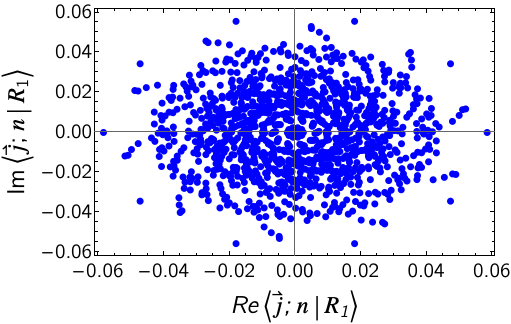}
\caption{Plot of the real and imaginary parts of the amplitudes
$\langle\vec{j};m|R_n\rangle$ for randomly selected other right
  eigenstates $|R_n\rangle$.}
\label{fig:allN5}
\end{figure}

%\subsection{Momentum $P=-2\pi/L$ eigenstates for $L=2N=10$}
\label{ssec:N5}
%%%%%%%%%%%%%%%%%%%%%%%%%%%%%%%%%%%%%%%%%%%%%%%%%
In the $P=2\pi/L$ momentum sector, there are a total of
\beq
\frac{1}{2}\begin{pmatrix}
  L\\
  L/2
\end{pmatrix}
\eeq
eigenstates $\langle L_n|$ of the transition matrix. The particular
state $\langle L_1|$ belonging to the State 3 family of Bethe states corresponds
to the following solution of the Bethe equations (for the left
eigenvectors)
\begin{itemize}
\item{} $L=2N=10$, $\ln(E)=-0.213449+0.702038i$
\begin{align}
u_1&=0.427136-0.873870i ,\
u_2=0.564221+0.680842i ,\
u_3=1.33637+0.597608i,\nn
u_4&=1.35093-0.646121i ,\ u_5=1.66292-0.005111i\ .
\label{rootsN5}
\end{align}
\item{} $L=2N=12$, $\ln(E)=-0.145070+ 0.573101i$
\begin{align}
u_1&=0.421706 - 0.736742 i ,\ u_2=0.451233 + 0.565759 i ,\
u_3=1.11218 + 0.639909 i ,\nn
u_4&=1.18183 - 0.655295 i ,\ u_5=1.52658 + 0.28296 i ,\ u_6=1.55750 - 0.23499 i\ .
\label{rootsN6}
\end{align}
\item{} $L=2N=14$, $\ln(E)=-0.105168 + 0.484332 i$
\begin{align}
u_1&=0.389443 + 0.481886 i ,\ u_2=0.408221 - 0.644572 i ,\
u_3=0.963278 + 0.632744 i ,\nn
u_4&=1.05742 - 0.647857 i ,\ u_5=1.38015 + 0.42379 i ,\ u_6=1.43968
- 0.36951 i\ ,\nn
u_7&=1.55081 + 0.03846 i\ .
\label{rootsN7}
\end{align}
\end{itemize}
In order to gain insight into the structure of all eigenstates in the
$P=2\pi/L$ momentum sector, we have determined their expressions in
both the configuration basis \fr{basisstates} and the momentum
eigenbasis \fr{mtmEVs} by numerically computing the overlaps 
\beq
\langle\vec{j};m|R_n\rangle\ ,\qquad
\langle \frac{2\pi}{L};\vec{j}_0|R_n\rangle\ .
\eeq
We first consider the amplitudes in the configuration basis. The state
\fr{rootsN5} has a set of amplitudes shown in Fig.~\ref{fig:N5}.
We see that in the product basis $\{\langle\vec{j};m|\}$ the
distribution of angles is roughly uniform, while the distribution of
magnitudes is strongly peaked around $0.028$. 
\begin{figure}[ht]
\centering
\includegraphics[width=0.45\linewidth]{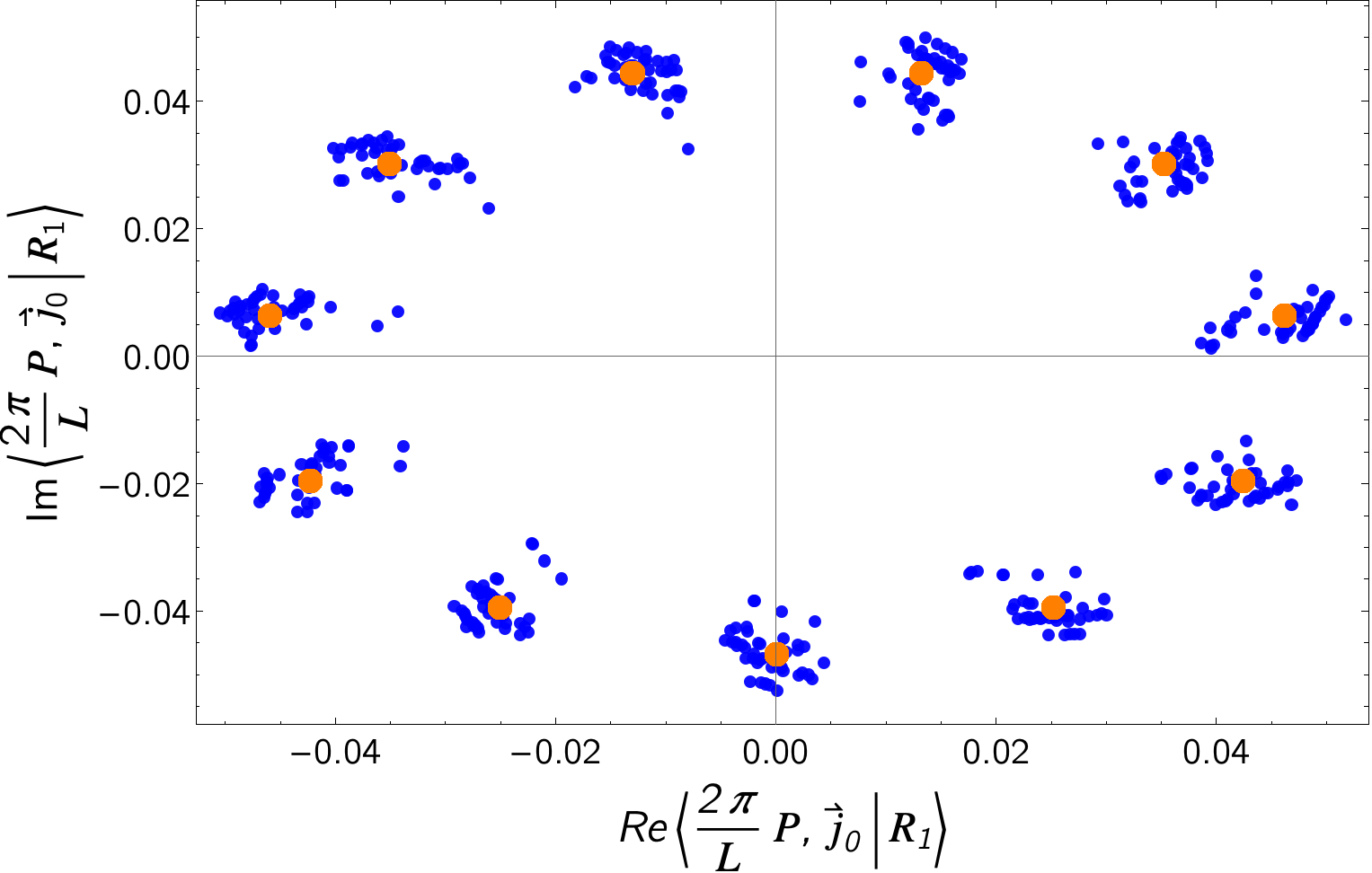}\quad
\includegraphics[width=0.45\linewidth]{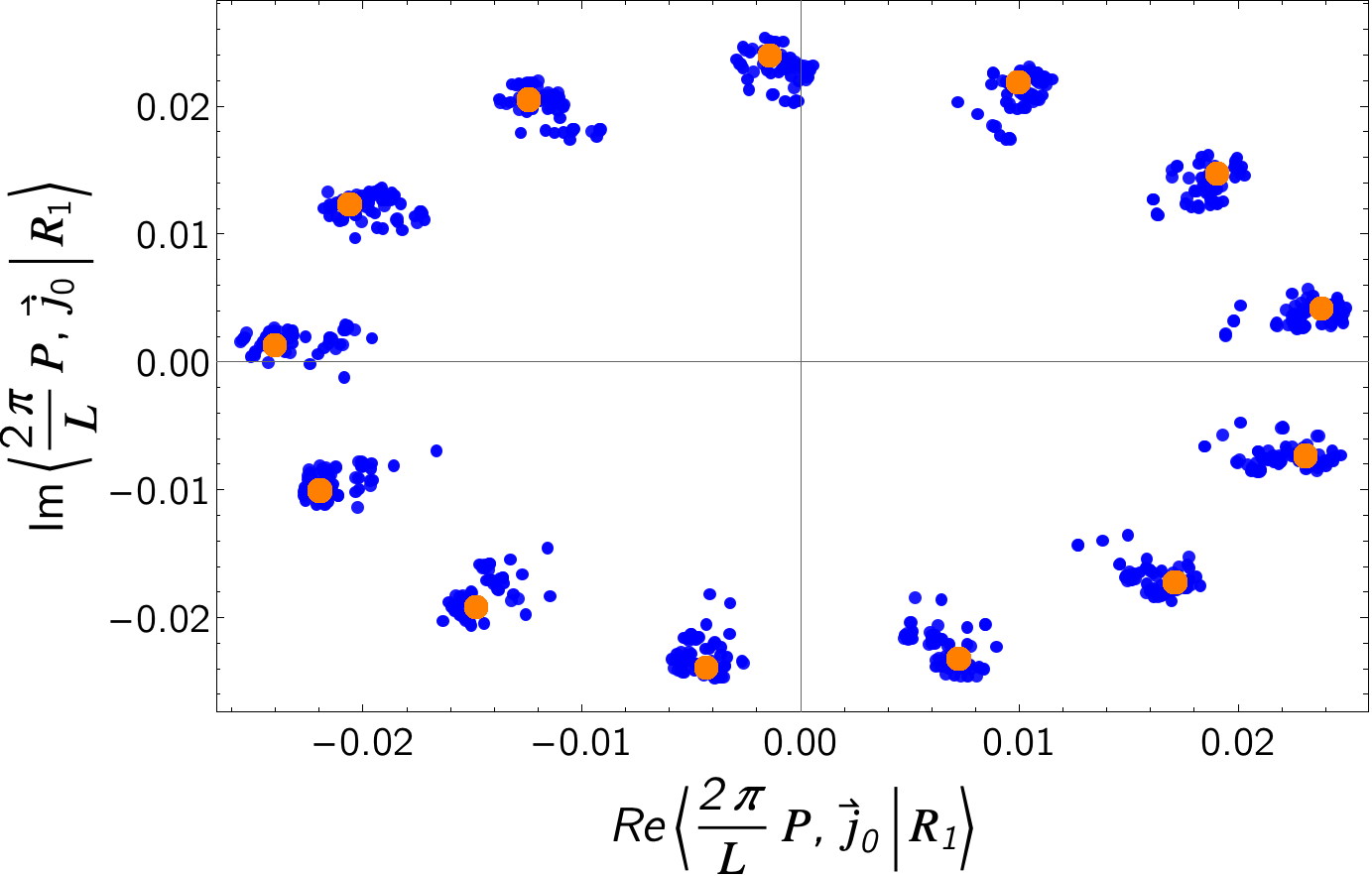}
\caption{Real vs imaginary parts of the amplitudes
$\langle\frac{2\pi}{L};\vec{j}_0|R_1\rangle$ for $N=6$ (left panel)
  and $N=7$ (right panel), where $|R_1\rangle$ are the eigenstates
\fr{rootsN6} and \fr{rootsN7}. The orange dots are the amplitudes
$\langle\frac{2\pi}{L};\vec{j}_0|F_{1,\psi}\rangle$ for $\psi=\frac{5\pi i}{22}$ and
$\psi=\frac{3\pi i}{52}$ respectively. }
\label{fig:N67}
\end{figure}
The other states have distributions of $\langle\vec{j};m|R_n\rangle$
that look qualitatively different. Examples are shown in \fig{fig:allN5}.
%%%%%%%%%%%%%%%%%%%%%%%%%%%%%%%%%%%%%%%%%%%%%%%%%
\subsubsection{Approximate structure of State 3}
%%%%%%%%%%%%%%%%%%%%%%%%%%%%%%%%%%%%%%%%%%%%%%%%%
In order to understand the physical properties of State 3, we propose
an approximate variational description of the form
\beq
|F_{r,\psi}\rangle=e^{i\psi}\sum_{\vec{j}_0}r_{\vec{j}_0}
%r_{\vec{j}_0}
e^{i\phi_{\vec{j}_0}}\ |\frac{2\pi}{L};\vec{j}_0\rangle \ ,
\label{FVector}
\eeq
where we choose
%\begin{align}
%\phi_{\vec{j}_0}&=\frac{4\pi X_{\vec{j}_0}}{L-1}\ ,\quad
%r_{\vec{j}_0}=(1+\epsilon)-2\epsilon\frac{X_{\vec{j}_0}}{X_{\rm
%    max}}\ ,\nn
%X_{\vec{j}_0}&=\sum_{m=2}^{L/2}\big(j_{0,m}-\frac{L+2}{2}\big)\ ,\qquad
%X_{\rm max}={\rm max}_{\vec{j}_0}X_{\vec{j}_0}\ .
%\end{align}
\begin{align}
\phi_{\vec{j}_0}&=\frac{4\pi X_{\vec{j}_0}}{L-1}\ ,\quad
X_{\vec{j}_0}=\sum_{m=2}^{L/2}\big(j_{0,m}-\frac{L+2}{2}\big)\ .
\end{align}
Here, $X_{\vec{j}_0}$ is obtained by removing the first site (pointer
position) from the configuration $\{\vec{j}_0;1\}$, and then summing
over the distances of the remaining particles from the center of the
lattice.
%\begin{equation}
%\{\vec{j},L-1\}=\overset{1}{\ONE} \ONE \ZERO\ZERO\ONE\ZERO\THREE
%\overset{L}{\ZERO}\rightarrow
%\{\vec{j}_0,1\}=\overset{1}{\THREE} \ZERO \ONE\ONE\ZERO\ZERO\ONE \overset{L}{\ZERO}
%X_{\vec{j}_0}=\sum_{m=2}^{L/2}\big(j_{0,m}-\frac{L+2}{2}\big)\ .
%\end{equation}
In the configuration basis, the
state reads
\beq
|F_{r,\psi}\rangle=e^{i\psi}\sum_{\{\vec{j};n\}}
r_{\vec{j}_0}
e^{i\phi_{\vec{j}_0}+\frac{2\pi
    ij_n}{L}}\ |\vec{j};n\rangle\ ,\ \quad\text{ where }
|\vec{j};n\rangle=\hat{\tau}^{n-1}|\vec{j}_0;1\rangle\ .
\eeq
The amplitudes of the simplest variational state $|F_{1,\psi}\rangle$
are shown in \fig{fig:N5} and \fig{fig:N67} for $L=2N=10,12,14$,
respectively, and are seen to approximately track those of State 3.

%%%%%%%%%%%%%%%%%%%%%%%%%%%%%%%%%%%%%%%%%%%%%%%%%
\subsubsection{Overlaps involving State 3}
%%%%%%%%%%%%%%%%%%%%%%%%%%%%%%%%%%%%%%%%%%%%%%%%%
We now turn to the overlaps
\beq
\omega_N(Q)\equiv\langle u|\hat{S}(-Q)|R_1\rangle\langle L_1|\hat{S}(Q)|P_{\rm SS}\rangle\ ,
\eeq
where
\beq
\hat{S}(Q)|\vec{j},n\rangle=\frac{1}{\sqrt{L}}\sum_{k=1}^Ne^{iQj_k}|\vec{j},n\rangle\ .
\eeq
By momentum conservation this vanishes unless $Q=2\pi/L$. Numerically,
we find for small systems

\beq
\omega_5\glb\frac{2\pi}{L}\grb=2.69504\times 10^{-6}\ ,\
\omega_6\glb\frac{2\pi}{L}\grb=1.71279\times 10^{-7}\ ,\
\omega_7\glb\frac{2\pi}{L}\grb=1.11663\times 10^{-8}.
\eeq
These are compatible with an exponential decay in $N$, and their
smallness provides an explanation of why the gap associated with State 3
is not detected in MC simulations. 
%%%%%%%%%%%%%%%%%%%%%%%
\subsection{Dynamical response in the \LT}
%%%%%%%%%%%%%%%%%%%%%%%
Our analysis is compatible with the following scenario, which is
premised on the assumption that there is at most a polynomial number
number of eigenstates of the transition matrix with eigenvalues such
that ${\rm   Re}\ln(E_n)\sim L^{-2}$, and their matrix
elements are exponentially small in $L$:
\beq
\chi_{{\cal AB}}(t)\sim
\begin{cases}
c_1 e^{-c_2 t/L^{3/2}} & \text{if } 1\ll t\sim L^{3/2}\ ,\\
c_3 e^{-\gamma L} e^{-c_4 t/L^2} & \text{if }  L^{5/2}\lesssim t\ .
\end{cases}
\eeq
Here $c_j={\cal O}(1)$. In this scenario the eventual exponential
decay of the susceptibility with the relaxation time (inverse spectral
gap) is not detectable by numerical methods because the susceptibility
is too small to be reliably computed in the relevant time window.

\begin{figure}[ht]
\centering
\includegraphics[width=0.5\linewidth]{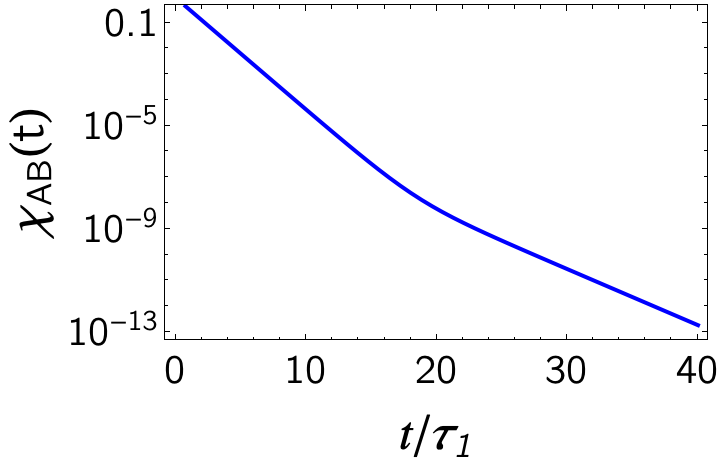}
\caption{Cartoon of proposed behaviour of dynamical response functions
$\chi_{\cal AB}(t)$ in equilibrium of the \LT. A exponential decay
with rate $\tau_1^{-1}\sim L^{-3/2}$ over a large intermediate time window is
followed by the true asymptotic decay with rate $\tau_2^{-1}\sim
L^2$. The latter regime is undetectable numerically because the
susceptibility is already extremely small.
}\label{fig:xover}
\end{figure}

%%%%%%%%%%%%%%%%%%%%%%%%%%%%%%%%%%%%%%%%%%%%%%%%%
\section{Continuum limit of the \LT}
\label{sec:contlim}
%%%%%%%%%%%%%%%%%%%%%%%%%%%%%%%%%%%%%%%%%%%%%%%%%
In order to facilitate the continuum limit, it is useful to introduce a
step-size $\Delta t$ for the time evolution and rewrite the
master equation as
\begin{align}
P(t+\Delta t)&=TP(t)=P(t)+\Delta t\ \frac{T-1}{\Delta t}P(t)\ .
\end{align}
Rearranging then gives
\beq
\frac{P(t+\Delta t)-P(t)}{\Delta t}=MP(t)\ ,\qquad M=\frac{T-1}{\Delta t}.
\eeq
The eigenvectors of $M$ are the same as the ones of $T$, and the
eigenvalues are related by
\beq
{\cal E}=\frac{E-1}{\Delta t}\ .
\eeq

%%%%%%%%%%%%%%%%%%%%%%%%%%%%%%%%
\subsection{Taking the continuum limit}
%%%%%%%%%%%%%%%%%%%%%%%%%%%%%%%%
\label{sec:TakingContinuum}
We introduce a lattice spacing $a_0$, such that the sites of our
$L$-site lattice are at positions $ja_0$. The physical length of the lattice
is then
\beq
\mathfrak{L}=La_0\ .
\label{Fequ:frakLDefinition}
\eeq
The continuum limit corresponds to taking $a_0\to 0$, $L\to\infty$,
while keeping $\mathfrak{L}$ fixed. In the continuum
limit it is useful to rewrite \eq{ansatz} for the
amplitudes. Introducing continuous particle positions as
\beq
x_{j_n}=a_0j_n\ ,\quad n=1,\dots, N
\eeq
we have
\beq
\psi_a(\boldsymbol{x})=\sum_{Q\in S_N}A_a(Q)\prod_{j=1}^N
e^{\frac{\ln(z_j)}{a_0}x_{Q_j}}\ .
\eeq
In order to retain a non-trivial dependence on the particle positions $x_j$
we introduce rescaled rapidity variables $u_j$ by
\beq
\ln(z_j)=u_ja_0\ .
\eeq
This results in
\begin{align}
\prod_{j=1}^N\left[E-\frac{1-\alpha}{z_j} \right]&=
\prod_{j=1}^N\left(E-(1-\alpha)+a_0(1-\alpha)u_j \right)\ ,\nn
\prod_{k=1}^N\frac{\alpha}{z_k}&=\alpha^Ne^{-a_0\sum_{k=1}^Nu_k}=\alpha^N\big[1-
a_0\sum_{k=1}^Nu_k+{\cal O}(a_0^2)\big]\ ,
\end{align}
which in turn suggests the expansion
\beq
E=1+a_0\epsilon\ .
\label{E}
\eeq
In order to obtain a non-trivial scaling limit we recall that the
eigenvalues of $M$ are related to $E$ by
\beq
{\cal E}=\frac{E-1}{\Delta t}=\frac{a_0}{\Delta t}\epsilon.
\eeq
This means that the continuum limit requires going over to a
continuous-time description
\beq
\Delta t, a_0\to 0\ ,\qquad \frac{a_0}{\Delta t}=v=\text{ fixed}.
\label{equ:ContinuousTime}
\eeq
Here $v$ is a characteristic propagation velocity. Finally, we rescale
the pullback parameter $\alpha$ as 
\beq
\alpha=a_0\mathfrak{a}\ .
\eeq
The Bethe equations then become
\begin{align}
e^{\mathfrak{L}u_j}&=(-1)^{N+1}\left(1+\frac{\epsilon+u_j}
{\mathfrak{a}}\right)\prod_{b=1}^N\frac{\epsilon+u_b}{\epsilon+u_j}\
,\nn
1&=\prod_{j=1}^N\Big(1+\frac{\epsilon+u_j}{\mathfrak{a}}\Big)\ .
\end{align}
We may change variables to bring the Bethe equations into a nicer form
\beq
w_j=\frac{\epsilon+u_j}{\mathfrak{a}}\ .
\eeq
Then

\begin{align}
e^{\mathfrak{L}(\mathfrak{a}w_a-\epsilon)}&=(-1)^{N+1}(1+w_a)\prod_{b=1}^N\frac{w_b}{w_a}\
,\nn
1&=\prod_{j=1}^N(1+w_j)\ .
\end{align}

%%%%%%%%%%%%%%%%%%%%%%%%%%%%%%%%
\subsubsection{One and two-particle sectors}
%%%%%%%%%%%%%%%%%%%%%%%%%%%%%%%%
For $N=1$ we have $w_1=0$, which translates into
\beq
\epsilon=u_1=\frac{2\pi i n}{\mathfrak{L}}\ ,\quad n\in\mathbb{Z}.
\eeq
In the two-particle sector $N=2$ the Bethe equations can be reduced to
a simple quadratic equation
\beq
\epsilon=\mathfrak{a}\frac{w_1+w_2}{2}\text{ mod} \frac{2\pi i}{\mathfrak{L}}
\ ,\qquad
w_2=-\frac{w_1}{1+w_1}\ ,\qquad
w_1^2+2w_1\big(1+\frac{\pi i n}{\mathfrak{aL}}\big)+\frac{2\pi i n}{\mathfrak{aL}}=0.
\eeq

%%%%%%%%%%%%%%%%%%%%%%%%%%%%%%%%
\subsection{Master equation in the continuum limit}
%%%%%%%%%%%%%%%%%%%%%%%%%%%%%%%%
In the continuum limit configurations are labelled by $N$ co-ordinates
$x_j$ and the position $a$ of the pointer among the particles. Let us denote
the probability distribution function corresponding to a configuration
$(\boldsymbol{x},a)$ by
\beq
P_a(\boldsymbol{x})\ .
\eeq
By taking the ``naive'' continuum limit of the discrete master equation
we obtain
\beq
\frac{1}{v}\frac{\partial P_{a}(\boldsymbol{x};t)}{\partial t}=
-\frac{\partial}{\partial
 x_a}P_a(\boldsymbol{x};t)+\mathfrak{a}[P_{a+1}(\boldsymbol{x};t)-P_{a}(\boldsymbol{x};t)]+\delta(x_a-x_{a-1})\big[P_{a-1}(\boldsymbol{x};t)-P_{a}(\boldsymbol{x};t)\big]\ .
\label{CME}
\eeq
Some of the relevant steps are
\begin{enumerate}
\item{} On the lattice the probability of a configuration
$(\boldsymbol{j},a)$ is $\psi_a(\boldsymbol{j})$. In the continuum limit the
particle positions are given by
\beq
x_n=j_na_0\ .
\eeq
\item{} The discrete time master equation on the lattice is
\begin{align}
\psi_a(\boldsymbol{j};t+\Delta t)&=
\bar{\alpha} \psi_{a-1}(\boldsymbol{j};t)\ \delta_{j_{a},j_{a-1}+1}
+\bar{\alpha}\psi_a(\dots,j_a-1,\dots;t)\
(1-\delta_{j_a,j_{a-1}+1})\nn
&+\alpha
\psi_{a+1}(\dots,j_{a+1}-1,\dots;t)\ (1-\delta_{j_{a+1},j_{a}+1})
+\alpha \psi_{a}(\boldsymbol{j};t)\ \delta_{j_{a+1},j_{a}+1}\ .
\label{SG}
\end{align}
%\item{}
Setting
\beq
P_a(\boldsymbol{x};t)=\psi_a(\boldsymbol{j};t)\ ,
\eeq
and using that (this identity is understood in terms of a
summation/integration over a test function)
\beq
\delta_{j,k}\rightarrow a_0\delta(x-x')
\eeq
we obtain
\begin{align}
P_a(\boldsymbol{x};t+\Delta t)&=
(1-a_0\mathfrak{a}) P_{a-1}(\boldsymbol{x};t)\
a_0\delta(x_a-x_{a-1}-a_0)\nn
&+(1-a_0\mathfrak{a}) P_{a}(\dots x_a-a_0,\dots;t)\ \big(1-a_0\delta(x_a-x_{a-1}-a_0)\big)\nn
&+\mathfrak{a}
P_{a+1}(\dots,x_{a+1}-a_0,\dots;t)\ \big(1-a_0\delta(x_{a+1}-x_{a}-a_0)\big)\nn
&+\mathfrak{a} P_{a}(\boldsymbol{x};t)\ a_0\delta(x_{a+1}-x_a-a_0)\ .
\label{SG2}
\end{align}
\item{} Finally, we expand in $a_0$ and drop all terms of order $a_0^2$ to
arrive at \eq{CME}.
\end{enumerate}

%%%%%%%%%%%%%%%%%%%%%%%%%%%%%%%%%%%%%%%%%%%%%%%%%%%%%%%%%%%%%%%%%%%
\subsection{Markov-process interpretation of the continuum limit}
\label{sect:MarkovProcess}
%%%%%%%%%%%%%%%%%%%%%%%%%%%%%%%%%%%%%%%%%%%%%%%%%%%%%%%%%%%%%%%%%%%
As introduced in \sect{sec:TakingContinuum}, we consider the continuum limit of
the \LT of $N$ particles on $L$ lattice sites, where $L \to \infty$.
In this limit, it samples the partition function of a gas of hard spheres with
vanishing radius $\sigma \to 0$. In the original model with $L$ sites, and
in terms of the original scales of length and time (see
\REF[Sec. III.A]{essler2024lifted}), the mean activiy drift per single
step is given by  
\begin{equation}
\mean{v_{\rightarrow}}= -\alpha \frac{L}{N} + 1\ .
\end{equation}
Introducing the lattice constant $a_0$ and the continuous time with a velocity
$v=1$ as in \eqtwo{Fequ:frakLDefinition}{equ:ContinuousTime}, this yields
\begin{equation}
\mean{v_{\rightarrow}} = -\afrak \frac{\Lfrak}{N} + 1\ .
\end{equation}
In the continuum limit, the \LT becomes equivalent to the factor-field
event-chain Monte Carlo algorithm  for zero-diameter one-dimensional hard
spheres~\cite{Lei2019}, and the Markov chain governing the \LT turns into a
continuous-time Markov process.
The active sphere moves with positive velocity $v=1$ either until it collides
with its right-hand neighbour or until a Poisson clock of intensity $\afrak$
moves the pointer to its left-hand neighbour. The collision between
zero-diameter objects is of course somewhat artificial as, at the collision,
the particle coordinates are the same.

For event-chain Monte Carlo, the pointer drift is an estimator for the system
pressure~\cite{Michel2014JCP} and indeed, for $\afrak = 0$, we recover the
well-known formula
\begin{equation}
 \beta P = \frac{\partial }{\partial \Lfrak} Z(N, \Lfrak, \sigma)=
 \frac{N}{\Lfrak} \mean{v_{\rightarrow}} =  \frac{N}{\Lfrak},
\end{equation}
where the hard-disk partition function is $ Z(N, \Lcont) = \Lcont^N$,
in other words the partition function of one-dimensional hard spheres (for
$\sigma =0$) first computed by Tonks in 1936~\cite{Tonks1936,SMAC}.

For finite $\afrak$, the
partition function is of a modified hard-sphere model
\begin{equation}
U = \sum_{k,l} U^{\text{hs}}(x_k, x_l) + \afrak \sum_{k=1}^N (x_k -
x_{k-1})\ ,\quad x_{N+1}\equiv x_1 + \Lfrak.
\label{equ:FactorFieldEnergy}
\end{equation}
Here, in addition to the hard-sphere potential $U^{\text{hs}}$, there are
linear attractions of strength $\afrak$ between nearest neighbours
called factor fields, which do not modify the particle statistics, as they
sum up to $\afrak \Lfrak$ (see \REF{Krauth2024hamiltonian} for a discussion).
Even for the motion of a single particle $x_k$,
the factor-field term in \eq{equ:FactorFieldEnergy} remains unchanged.
Nevertheless, it influences the \LUT dynamics, in which four factors
independently influence the motion, namely the factor field of $x_k$ with its
forward neighbour (positioned at $x_{k+1}$, up to boundary conditions), the
backward neighbour (at $x_{k-1}$), and identically for the hard-sphere
interaction (see \sect{sec:FactorizedMetropolisFilter} for a discussion of this
point in the language of the factorized Metropolis algorithm). The weight of
each configuration is now
\begin{equation}
\pi(\SET{x_1 \TO x_N}) = \expb{-\beta \afrak \Lfrak}.
\end{equation}
The partition function is:
\begin{align}
 Z_\afrak(N, \Lcont) &= \Lcont  ^N \expa{-\beta \Lfrak \afrak},
 \label{equ:PartitionFunctionFactor}
\end{align}
with the pressure (for $\beta = 1$):
\begin{equation}
P_\afrak(N, \Lfrak) = \partial \logc{Z_\afrak (N, \Lfrak)} /
\partial \Lfrak |_{\beta = 1} = N /  \Lcont - \afrak.
\end{equation}
Multiplied with $\Lfrak / N$, the pressure trivially equals the pointer drift,
a finding that holds much beyond the trivial $\sigma=0$ case treated
here, and that remains valid in more than one
dimension~\cite{Michel2014JCP,Li2022}.

The $\bigOb{N^{3/2}}$ scaling for the autocorrelation times of the structure
factor has been observed in the hard-sphere model also at finite $\sigma$ at
the critical factor field, which becomes, for large $N$, $\afrak = N / (\Lfrak
- N \sigma) $~\cite{Lei2019}. In many other one-dimensional models, such as the
Lennard-Jones model or the harmonic chain~\cite{Krauth2024hamiltonian}, this
optimal scaling was found for a critical factor field which again corresponds to
a vanishing pointer drift and, thus, to vanishing pressure.

%%%%%%%%%%%%%%%%%%%%%%%%%%%%%%%%%%%%%%%%%%%%%%%%%%%

\section{Generalized Lifted TASEP}
\label{sec:GeneralizedLiftedTASEP}

In the present section, we generalize the \LT beyond hard-sphere interactions,
but retain that, at each time step, a uniquely defined active particle moves
in the forward direction. In the continuum, this generalization carries over to
arbitrary continuous pair interactions but also to many-body potentials. It is
the factorized Metropolis algorithm~\cite{Michel2014JCP}. On the lattice, only
monotonous interactions between neighbouring particles allow for a generalization
of the \LT. For these cases, we can set up a lifted algorithm which conserves
the Boltzmann distribution as a stationary distribution.

\subsection{Factorized Metropolis filter}
\label{sec:FactorizedMetropolisFilter}

In the \GLT, configurations do not all have the same statistical weight, but
rather  Boltzmann weights given by the nearest-neighbour
interactions that can be written as a potential:
\begin{equation}
\expb{-\beta U_{\dots, i,j,k, \dots}} =
\pi_{\dots, i,j,k,\dots} =  \cdots \pi_{j-i} \pi_{k-j} \cdots.
\end{equation}
Here,
$\dots, i, j, k, \dots$ are particle positions and the
nearest-neighbour  pairs, such as $(i,j)$ and $(j,k)$, are referred to as
\quot{factors}. Periodic boundary conditions are understood. In the factorized
Metropolis filter~\cite{Michel2014JCP}, each factor accepts the move
individually with its
Metropolis filter. A move of the active particle  $j \to j+1$ is thus accepted
with probability
\begin{equation}
\pfact(j \to j+1) =
 \minb{1, \frac{\pi_{j+1-i} }{\pi_{j-i}}}
 \minc{1, \frac{\pi_{k- (j+1)} }{\pi_{k-j}}}\ .
 \label{equ:FactorizedSimplified}
\end{equation}
We have to  avoid that both factors in \eq{equ:FactorizedSimplified} reject the
move simultaneously, in which case the pointer cannot be re-attributed.
We thus require the interaction to be monotonous and, for concreteness,
repulsive:
\begin{equation}
 \pi_k \ge \pi_l\quad \text{for $k> l$}.
\end{equation}
In the factorized Metropolis filter of \eq{equ:FactorizedSimplified}, the
factor $(i,j)$ always accepts the (forward) move of particle $j$, so that
\eq{equ:FactorizedSimplified} further simplifies to:
\begin{equation}
\pfact(j \to j+1) =
\minc{1, \frac{\pi_{k- (j+1)} }{\pi_{k-j}}} = \frac{\pi_{k- (j+1)} }{\pi_{k-j}}
= : p_{k-(j+1)}\ .
 \label{equ:FactorizedForward}
\end{equation}
The \LT of \eqtwo{equ:LTASEP1}{equ:LTASEP2} corresponds to the choice
$\SET{\pi_0, \pi_1, \pi_2\dots} = \SET{0,1,1, \dots}$

The move from forward distance $4$ toward $3$ is accepted with
probability $p_3$, toward $2$ with probability $p_2$, and toward
$0$ with probability $p_0$. In the hard-sphere case, $p_0 = 0$.
The simplest generalization is
when   $0< p_1< 1$ and $p_k = 1$ for $k > 1$. In this case,  we call
$p: = p_1$.

\subsubsection{Definition of the \GLT}
As in the \TASEP, we consider the evolution probabilities for a  lifted
configuration
\begin{equation}
 \overset{i}{\ONE}  \overset{\vv{\jmath}}{\TWO} \ZERO
\overset{k}{\ONE}.
\end{equation}
Again, a single particle $j$ is active (carries the pointer), but its advance
can be rejected at a distance by its forward neighbour $k$. The backward
neighbour does not reject the move under the factorized Metropolis algorithm, as
the potential is supposed repulsive.
In the first half-step of the move, the displacement of the active
particle from $j $ to $j+1$ is accepted  with
the Metropolis ratio of the weights $\pi_{k -
(j+1)}/\pi_{k-j}$ (see \eq{equ:FactorizedForward}).
If the move is rejected, the particle at $k$ obtains the pointer.
The second half-step of the move performs a pullback move with
probability $\alpha$. In total:
\begin{equation}
\overset{i}{\ONE}  \overset{\vv{\jmath}}{\TWO} \ZERO
\overset{k}{\ONE}
\rightarrow
\begin{cases}
 \overset{i}{\ONE} \ZERO \TWO \ONE\quad p_{k-(j+1)} &
\rightarrow
\begin{cases}
 \TWO \ZERO \ONE \ONE \quad p_{k-(j+1)} \alpha \\
 \ONE \ZERO \TWO \ONE \quad p_{k-(j+1)} \alphabar
\end{cases}
\\
 \ONE \ONE \ZERO \TWO \quad \pbar_{k-(j+1)}&
\rightarrow
\begin{cases}
 \ONE \TWO \ZERO \ONE \quad \pbar_{k-(j+1)} \alpha \\
 \ONE \ONE \ZERO \TWO \quad \pbar_{k-(j+1)} \alphabar,
\end{cases}
\end{cases}
\label{equ:RowSumsTransitionMatrix}
\end{equation}
(Here, we set $\alphabar = 1 - \alpha$ and $\pbar = 1 - p$.)
The system in \eq{equ:RowSumsTransitionMatrix} requires only the
positions of the three particles $i, \vv{\jmath},k$ in order to define the
evolution of an arbitrary
large system with periodic boundary conditions, in which $\vv{\jmath}$ is
active.
\eqq{equ:RowSumsTransitionMatrix} shows that the rows
of the transition matrix
sum up to one, in other words that $(1 \TO 1)$ is a right eigenvector of $P$
with eigenvalue $1$. The right-eigenvector equation corresponding to
\eq{equ:RowSumsTransitionMatrix} is:
\begin{align}
E \rightEV{i, \vv{\jmath}, k} &=
p_{k-(j+1)}\alpha\  \rightEV{\vv{\imath}, j+1, k} +
p_{k - (j+1)}\alphabar\  \rightEV{i, \vv{\jmath+1}, k}\nn
&+\glc 1-p_{k-(j+1)} \grc\alpha\  \rightEV{i, \vv{\jmath}, k} +
\glc 1-p_{k-(j+1)} \grc \alphabar\  \rightEV{i, j+1, \vv{k}},
\label{equ:RightEigenvectors}
\end{align}
where only one of the $p_k$  appears in all four terms.
For any choice of the parameters $p_1, p_2, \dots$ in the   \GLT, we have
three sets of such closed equations for the right eigenvectors.

\subsection{Global balance for the \GLT}

To establish the global-balance condition, we
use that the factorized Metropolis algorithm, although
it is
used exclusively in a non-reversible setting, nevertheless satisfies the
detailed-balance condition. The following example illustrates this for three
particles $(i,\vv{\jmath},k)$.
\begin{equation}
\begin{rcases*}
\begin{rcases*}
 \overset{i}{\ONE} \overset{\!\!\! \vv{\jmath-1} \!\!\!}{\TWO} \ZERO
\ZERO \overset{k}{\ONE}\ p_{k-j}
\\
 \TWO \ZERO \ONE \ZERO \ONE\ 1 - p_{j-(i+1)}
\end{rcases*}
\rightarrow
 \overset{i}{\ONE} \ZERO \overset{\vv{\jmath}}{\TWO} \ZERO
\overset{k}{\ONE} & $p = \alphabar$\\
\begin{rcases*}
 \ONE \ZERO \ONE\overset{\!\!k-1\!\!}{\TWO} \ZERO\ p_{l-k}   \\
 \ONE \ZERO \underset{j}{\TWO} \ZERO \ONE \ 1-p_{k-(j+1)}
\end{rcases*}
\rightarrow
 \overset{i}{\ONE} \ZERO \overset{j}{\ONE} \ZERO
\overset{\vv{k}}{\TWO} &
$p = \alpha$
\end{rcases*} \rightarrow
 \overset{i}{\ONE} \ZERO \overset{\vv{\jmath}}{\TWO} \ZERO
\overset{k}{\ONE}
\label{equ:GlobalBalanceGraphicsReduced}
\end{equation}
The coefficients of the $E=1$ eigenvector are the stationary probabilities, so
that the flow into configuration $(i,\vv{\jmath},k)$ is given by
\begin{align}
& \pi_{i,j-1,k}\pfact\glb i,\vv{\jmath-1},k \to i,\vv{\jmath}, k \grb \alphabar
\label{equ:GlobalBalanceEq1}\\
+& \pi_{i,j,k}\glc 1-\pfact\glb \vv{\imath},j,k \to \vv{\imath+1},j, k \grb \grc
\alphabar 
\label{equ:GlobalBalanceEq2}\\
+&\pi_{i, j, k-1} \pfact   \glb i,j, \vv{k-1}  \to i,j, \vv{k}    \grb \alpha 
\label{equ:GlobalBalanceEq3}\\
+&\pi_{i,j,k} \glc 1 - \pfact \glb i,\vv{\jmath}, k \to i, \vv{\jmath+1}, k
\grb     \grc \alpha\ .
\label{equ:GlobalBalanceEq4}
\end{align}
The detailed-balance condition allows one to turn around
\eqtwo{equ:GlobalBalanceEq1}{equ:GlobalBalanceEq3}, which gives
\begin{align}
& \pi_{i,j,k}\pfact\glb
(i,\vv{\jmath}), k \to (i,\vv{\jmath-1}),k
 \grb \alphabar
 \label{equ:GlobalBalanceEq1rev}\\
+ & \pi_{i,j,k}\glc 1-\pfact\glb (\vv{\imath},j),k \to (\vv{\imath+1},j), k
\grb
\grc \alphabar 
\label{equ:GlobalBalanceEq2rev}\\
+ & \pi_{i, j, k} \pfact   \glb
i,(j, \vv{k})    \to i,(j, \vv{k-1})
\grb \alpha 
\label{equ:GlobalBalanceEq3rev}\\
+ & \pi_{i,j,k} \glc 1 - \pfact \glb i,(\vv{\jmath}, k) \to i, (\vv{\jmath+1},
k)
\grb     \grc \alpha = \pi_{i,j,k}\ .
\label{equ:GlobalBalanceEq4rev}
\end{align}
In \eqtwo{equ:GlobalBalanceEq1rev}{equ:GlobalBalanceEq2rev}, as indicated, the
interval $(i,j)$
is shortened from either side (and the sum of these two terms gives
$\pi_{i,j,k} \alphabar$, and
likewise, in
\eqtwo{equ:GlobalBalanceEq3rev}{equ:GlobalBalanceEq4rev}, the interval $(j,k)$
is shortened from either side (with the sum equal to $\pi_{i,j,k} \alpha$). It
follows that the flow into $(i,\vv{\jmath},k)$ equals $\pi_{i,j,k}$.
Global balance is satisfied, and the \GLT has the stationary distribution
$\pi$.

\begin{figure}[ht]
\centering
\includegraphics[width=0.80\linewidth]{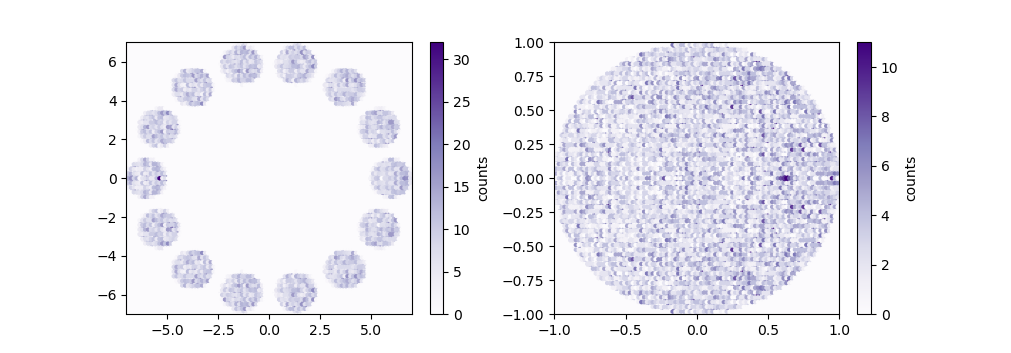}
\caption{Complex eigenvalue-ratio statistics for the \GLT.
\subcap{Left} Eigenvalue ratios separated into momentum sectors. No structure
is detected. \subcap{Right} Merged-momentum-sector eigenvalue ratios, which are
again structureless. The parameters are $N=7$, $L=14$, for $\alpha = 0.2$ and
$p=0.8$.}
\label{fig:CLS}
\end{figure}

To numerically test for the possible integrability, we checked for the
criterion proposed in \REF{Sa_Ribeiro_Prosen_2020} and computed complex ratios
of eigenvalues, both separated into momentum sectors and merged for all
momentum sectors (see \fig{fig:CLS}). The absence of structure
in the histogram points towards integrability, but a more careful
analysis for larger system sizes is required to have any confidence in this.
An investigation of the two and three-particle sectors did not reveal
any clear signs of integrability \cite{FE2025}.

%%%%%%%%%%%%%%%%%%%%%%%%%%%%
\section{Conclusions}
\label{sec:Conclusions}
%%%%%%%%%%%%%%%%%%%%%%%%%%%%
In this work we have studied the \LT and its generalization, the
\GLT. In the \LT we have carried our a large-scale numerical analysis
of the Bethe ansatz equations in order to show how the spectral gap
$\Delta$ crosses over, as a function of $\alpha$, between the
asymptotic $N^{-5/2}$ scaling of $\Delta$ at $\alpha\neq\alphacrit$
and the $N^{-2}$ scaling at $\alpha=\alphacrit$. An analogous
crossover is observed in Monte Carlo simulations of the integrated
autocorrelation time of the structure factor, but its scaling with
particle number is seen to follow a $N^{3/2}$ law.

We have provided a possible explanation for this discrepancy by
carefully analyzing, for small values of $N$, the properties of the
eigenvector of the transition matrix that gives rise to the $N^{-2}$
scaling for large particle numbers at $\alpha=\alphacrit$. We were
able to ``follow'' this family of eigenvectors in $N$ by using the
Bethe ansatz solution. We showed that for small particle
numbers the contribution of the eigenvector of interest to dynamical
susceptibilities is too small to be observed in Monte Carlo
simulations. We propose that this smallness of the relevant matrix
elements, combined with a small number of eigenvectors whose
eigenvalues scale as $L^{-2}$, makes it essentially impossible to
detect the asymptotic relaxation time numerically. To the best of our
knowledge this is an unusual situation in stochastic processes of many
interacting particles. However, such a scenario is not ruled out by
any mathematical theorems, and may occur much more generally.

Another key result obtained in our work is the construction of an
integrable continuum limit (in both space and time), which we related to
the hard-sphere event-chain Monte Carlo algorithm. This allowed us to
relate the pullback $\alpha$ is related to the pressure, and identify the
critical pullback $\alphacrit$ as corresponding to vanishing pressure,
\emph{cf.} \cite{Lei2019}.

Finally, we have generalized the \LT to a wide class of nearest-neighbour
interactions, which leads to lifted Markov chains with non-trivial
equilibrium steady states.

\section*{Acknowledgements}
This work was supported in part by the EPSRC under grant EP/X030881/1
(FHLE). FHLE thanks the Institut Henri Poincar\'e (UAR 839
CNRS-Sorbonne Universit\'e) and the LabEx CARMIN (ANR10-LABX-59-01) for their
support. JG thanks the Rudolf Peierls Centre for Theoretical Physics for
hospitality and financial support. 
Research of WK was supported by a grant from the Simons Foundation (Grant
839534, MET).

\bibliography{General,Gen_LTASEP}
\end{document}